\documentclass[seceq,preprint]{ptptex}
\usepackage{graphicx}

\newcommand{\wt}{\widetilde}
\newcommand{\wh}{\widehat}
\newcommand{\ol}{\overline}
\newcommand{\del}{\partial}
\newcommand{\ra}{\rightarrow}

\newcommand{\nn}{\nonumber}
\newcommand{\half}{\frac{1}{2}}
\def\tr{\mathop{\rm tr}\nolimits}

\newcommand{\vev}[1]{\left\langle #1 \right\rangle}
\newcommand{\bra}[1]{\langle #1 |}
\newcommand{\ket}[1]{| #1 \rangle}

\newcommand{\cF}{{\mathcal F}}
\newcommand{\cJ}{{\mathcal J}}
\newcommand{\cL}{{\mathcal L}}
\newcommand{\cV}{{\mathcal V}}
\newcommand{\cO}{{\mathcal O}}
\newcommand{\cA}{{\mathcal A}}
\newcommand{\cB}{{\mathcal B}}
\newcommand{\cM}{{\mathcal M}}
\newcommand{\bR}{\mathbb{R}}

\newcommand{\bZ}{\mathbb{Z}}
\newcommand{\ba}{{\boldsymbol{a}}}
\newcommand{\by}{{\boldsymbol{y}}}
\newcommand{\bk}{{\vec{k}}}
\newcommand{\bn}{{\vec{n}}}
\newcommand{\bp}{{\vec{p}\,}}
\newcommand{\bx}{{\vec{x}}}
\newcommand{\bX}{{\vec{X}}}

\newcommand{\bsig}{{\vec{\sigma}}}

\newcommand{\Mkk}{M_{\rm KK}}

\newcommand{\psl}{\not\! p}

\pubinfo{Vol.~120, No.~6, December 2008}
\notypesetlogo 
\title{Holographic Baryons}

\subtitle{Static Properties and Form Factors\\
~~~~~~~~~~ from Gauge/String Duality}  

\author{Koji \textsc{Hashimoto},$^1$ 
Tadakatsu \textsc{Sakai}$^2$ and Shigeki \textsc{Sugimoto}$^3$}

\inst{$^{1}$Theoretical Physics Laboratory, RIKEN, 
Saitama 351-0198, Japan \\ 
$^{2}$Department of Physics, Faculty of Sciences, Ibaraki University, \\
Mito 310-8512, Japan \\ 
$^{3}$Institute for the Physics and Mathematics of the Universe,\\
The University of Tokyo, Chiba 277-8568, Japan}

\recdate{July 24, 2008}%

\abst{In this paper, we study the properties of baryons by using 
a holographic dual of QCD on the basis of the D4/D8-brane
configuration, where baryons are described by a soliton.
We first determine the asymptotic behavior of
the soliton solution, which allows us to evaluate 
well-defined currents associated
with the $U(N_f)_L \times U(N_f)_R$ chiral symmetry.
Using the currents, we compute the static quantities
of baryons such as charge radii and magnetic moments, and perform
a quantitative comparison with experiments. It is emphasized that
not only the nucleon but also excited baryons,
such as $\Delta$, $N(1440)$, $N(1535)$, etc.,
can be analyzed systematically in this model.
We also investigate the form factors and find that
our form factors agree well with the results that
have been well established empirically.
Using the form factors, the effective baryon-baryon-meson
cubic coupling constants among their infinite towers
in the model can be determined.
Some physical implications following from these results are discussed.}

\begin{document}
\maketitle

\section{Introduction}

The gauge/string duality has opened up
new technology to analyze strongly coupled gauge theories.
Although it has not been proven yet, there is
a lot of highly nontrivial evidence suggesting the duality
between four-dimensional gauge theories
and string theory in 
higher-dimensional curved backgrounds, at least for the
cases with supersymmetry and conformal symmetry
\cite{Maldacena:1997re,Gubser:1998bc,Witten:1998qj}
(for a review, see Ref.~\citen{ADS/CFT}).
It is obviously important to see if this idea can be
applied to realistic QCD in many respects.
One of the advantages of QCD is that
we can use experimental data to check if the duality
really works. In general, the analysis of non-supersymmetric
strongly coupled gauge theories is very difficult,
and hence it is almost impossible to establish such duality.
However, for QCD, we can omit the complicated calculations
on the gauge theory side and simply compare the results
on the string theory side with the experimental
data to obtain nontrivial evidence of the duality.

If the gauge/string duality is really applicable to QCD,
it will provide us with new deep insight
into the theory of strong interaction
in both conceptual and practical terms.
It will tell us that both QCD and string theory in a higher-dimensional
curved background (holographic QCD)
can be a fundamental theory
of strong interaction at the same time.
Of course, in the high-energy
weakly coupled regime, QCD is a better description.
However, at least in the large-$N_c$ strongly coupled regime,
the string theory can be used as a powerful tool to
calculate various physical quantities.

In Refs.~\citen{SaSu1} and \citen{SaSu2}, it was proposed that
QCD with $N_f$ massless quarks is dual to
type IIA string theory with $N_f$-probe D8-branes
in the D4-brane background used in Ref.~\citen{Witten:D4}
at low energy. In this model, mesons appear
as open strings on the D8-branes and
baryons are expressed as D4-branes wrapped on the nontrivial
four-cycle of the background, as was anticipated
in Refs.~\citen{Gross-Ooguri} and \citen{WittenBaryon}
in the context of the AdS/CFT correspondence.
The effective action for the open strings
turned out to be a five-dimensional $U(N_f)$ 
Yang-Mills-Chern-Simons (YM-CS) theory in a curved background.
Via the equivalence between the wrapped D4-branes and the instantons in
the D8-brane world-volume gauge theory \cite{brane-within-brane},
the baryons were obtained as instanton configurations in the
Euclidean four-dimensional space in this five-dimensional gauge theory
\cite{SaSu1}
(see also Ref.~\citen{Son-Stephanov}).

The idea of realizing baryons as solitons
was pioneered by Skyrme in the early 1960s 
\cite{Skyrme}.
He started with an effective action for a pion with
a fourth-derivative term (Skyrme term) and constructed
a soliton solution (Skyrmion) in which
the $U(N_f)$-valued pion field carries a nontrivial
winding number interpreted as the baryon number.
In the model of holographic QCD \cite{SaSu1,SaSu2},
this construction is naturally realized.
It can be shown that the five-dimensional gauge theory
is equivalent to a theory of mesons including pions
as well as infinitely many vector and axial-vector mesons.
The pion sector is given by the Skyrme model and
the instanton configuration reduces to a Skyrmion,
realizing the old idea of Atiyah and Manton \cite{Atiyah-Manton}
who proposed expressing a Skyrmion using a gauge configuration
with a non-zero instanton number.

The Skyrme model was further
developed by Adkins et al.\cite{ANW},
who analyzed various static properties of the nucleon
and $\Delta$ by quantizing the Skyrmion.
It is natural to expect that the analysis can be improved
by applying their strategy to the holographic model of QCD,
since the model automatically contains the contribution from the
vector and axial-vector mesons such as $\rho$, $\omega$, $a_1$,
etc.\footnote{See Ref.~\citen{NaSuKo} for an attempt to include
the effect of the $\rho$ meson in the Skyrmion based on the
holographic QCD.}
In fact, the quantization of the instanton
representing a baryon was studied in Refs.~\citen{HSSY,Hong-Rho-Yee-Yi,PaYi}.
It was found in Ref.~\citen{HSSY} that the spectrum of the baryons obtained
in the model, including nucleons, $\Delta$, $N(1440)$, $N(1535)$, etc,
is qualitatively similar to that observed in nature,
although the prediction of the mass differences tended to be larger than
the observed differences. In Refs.~\citen{Hong-Rho-Yee-Yi} and
\citen{PaYi}, 
a five-dimensional fermionic field was introduced into the
five-dimensional gauge theory to incorporate the baryon degrees
of freedom, where the interaction terms between the two fields
were analyzed by quantizing the instanton.
Using this five-dimensional system, the interaction of
the baryons with the electromagnetic field and mesons was computed,
which was found to be in good agreement with the experimental data.

In this paper, we analyze the static properties of baryons,
such as the magnetic moments, charge radii, axial couplings, etc.
of nucleons, $\Delta$, $N(1440)$, and $N(1535)$, 
generalizing Refs.~\citen{ANW} and \citen{HSSY}. These quantities
are obtained by calculating
the currents corresponding to the chiral symmetry
that contain the information of the couplings to
the external electromagnetic gauge field.
As we will see, most of our results are closer to
the experimental values than the results given
in Ref.~\citen{ANW} for the Skyrme model.
For some quantities of the excited baryons, our results 
are predictions.
We also consider the form factors of the spin 1/2 baryons
and study
how they behave as functions of the momentum
transfer.

Note that there are some closely related
works that address the same subject.
As mentioned above, Refs.~\citen{Hong-Rho-Yee-Yi}
and \citen{PaYi}
analyze the properties of baryons using the five-dimensional
gauge theory with a five-dimensional fermion.
On the other hand, our analysis is based on the quantum mechanics
obtained through the usual procedure of quantizing solitons.
These two approaches should in principle give the same results,
and in fact we will find that some of our results
are consistent with those in the studies, although the relation is not
completely transparent.
The work of Hata et al.~\cite{Hata-Murata-Yamato}
is more directly related to ours. The main difference is the
definition of the currents.
As discussed in their paper, their currents are problematic,
since they are not gauge-invariant and have some ambiguities
in the definition.

The paper is organized as follows.
In \S 2, after briefly reviewing the main points
of the model and the construction of the baryon,
we calculate the currents corresponding to the chiral symmetry.
The applications are given in \S 3. We calculate
various physical quantities including
charge radii, magnetic moments, axial coupling, etc.,
and compare them with the experimental values for nucleons.
We also present these values for excited baryons as our predictions.
In \S 4, we compute the form factors
of the spin 1/2 baryons and study their properties
including cubic baryon-baryon-meson couplings.
Section 5 is devoted to a summary and discussion
with tables summarizing our results 
including our predictions for future experiments.
Two appendices summarize some technical details.

\section{Currents}
\label{seccurrents}

\subsection{The model}
\label{rev:model}
In this subsection, we give a brief review of 
Refs.~\citen{SaSu1}, \citen{SaSu2}, and \citen{HSSY} with an emphasis on
the construction of baryons.\footnote{
Our notation is mostly consistent with Ref.~\citen{HSSY}
except that we do not use the rescaled variables
defined in (3.9) of Ref.~\citen{HSSY}.}
Based on the idea of gauge/string duality,
it was proposed that the meson effective theory
is given by a five-dimensional  $U(N_f)$ 
YM-CS theory in a curved background.
The action of the model is
\begin{align}
&S=S_{\rm YM}+S_{\rm CS}\ ,\nn\\
&S_{\rm YM}=-\kappa
\int d^4 x dz\,\tr\left[\,
\half\,h(z){\cF}_{\mu\nu}^2+k(z){\cF}_{\mu z}^2
\right]\ ,~~
S_{\rm CS}=\frac{N_c}{24\pi^2}
\int_{M^4\times\bR}\omega_5({\cA})\ .
\label{model}
\end{align}
Here, $\mu,\nu=0,1,2,3$ are four-dimensional Lorentz indices, and $z$
is the coordinate of the fifth dimension. The quantity
${\cA}=\cA_\alpha dx^\alpha=\cA_\mu dx^\mu+\cA_z dz
~~(\alpha=0,1,2,3,z)$ is the five-dimensional
$U(N_f)$ gauge field and $\cF=\frac{1}{2}\cF_{\alpha\beta}
dx^\alpha\wedge dx^\beta=d\cA+i\cA\wedge\cA$ is
its field strength. The constant $\kappa$ is related to the 't~Hooft
coupling $\lambda$ and the number of colors $N_c$ as
\begin{equation}
\kappa=\frac{\lambda N_c}{216\pi^3}\equiv a\lambda N_c\ .
\label{kappa}
\end{equation}
Although it is not explicitly written in (\ref{model}),
the mass scale of the model is given by a parameter
 $\Mkk$, which is the only dimensionful parameter of the model.
In Refs.~\citen{SaSu1} and \citen{SaSu2}, these two parameters
are chosen as
\begin{align}
 \Mkk=949\mbox{ MeV} \ ,~~~
 \kappa=0.00745 \ ,
\label{Mkkkappa}
\end{align}
to fit the experimental values of the $\rho$ meson mass and
the pion decay constant $f_\pi\simeq 92.4~{\rm MeV}$.
In this paper, we mainly work with $\Mkk=1$ unit.
The functions $h(z)$ and $k(z)$ are given by
\begin{equation}
h(z)=(1+z^2)^{-1/3}\ ,\quad k(z)=1+z^2\ ,
\label{hk}
\end{equation}
and $\omega_5(\cA)$ is the CS 5-form
defined as\footnote{
Here we omit the symbol $`\wedge'$ for the
wedge products of  $\cA$ and $\cF$
(e.g. $\cA\cF^2=\cA\wedge \cF\wedge\cF$)
in the CS 5-form.
}
\begin{equation}
\omega_5({\cA})=\tr\left(
\cA \cF^2-\frac{i}{2}\cA^3\cF-\frac{1}{10}\cA^5
\right)\ .
\end{equation}
The action (\ref{model}) is obtained in Ref.~\citen{SaSu1}
as the effective action of $N_f$-probe
D8-branes placed in the D4-brane background studied in
Ref.~\citen{Witten:D4} and is thought to be an effective theory of
mesons in four-dimensional (large $N_c$) QCD with $N_f$ massless
quarks.

In this paper, we consider the $N_f=2$ case, and the $U(2)$ gauge
field $\cA$ is decomposed as
\begin{eqnarray}
\cA=A+\wh A\,\frac{{\bf 1}_2}{2}
=A^a\frac{\tau^a}{2}+\wh A\,\frac{{\bf 1}_2}{2}
=\sum_{C=0}^3 \cA^C\,\frac{\tau^C}{2}\ ,
\label{decom}
\end{eqnarray}
where $\tau^a$ ($a=1,2,3$) are Pauli matrices and $\tau^0={\bf 1}_2$
is a unit matrix of size 2.
Then, the equations of motion are
\begin{align}
&-\kappa\left(
h(z)\del_\nu\wh F^{\mu\nu}+\del_z(k(z)\wh F^{\mu z})
\right)
+\frac{N_c}{128\pi^2}\epsilon^{\mu \alpha_2\cdots \alpha_5}
\left(
F^a_{\alpha_2\alpha_3}F^a_{\alpha_4\alpha_5}
+\wh F_{\alpha_2\alpha_3}\wh F_{\alpha_4\alpha_5}
\right)
=0\ ,
\label{eom1}\\
&-\kappa\left(
h(z)D_\nu F^{\mu\nu}+D_z(k(z) F^{\mu z})
\right)^a
+\frac{N_c}{64\pi^2}\epsilon^{\mu \alpha_2\cdots \alpha_5}
F^a_{\alpha_2\alpha_3}\wh F_{\alpha_4\alpha_5}=0\ ,
\label{eom2}\\
&-\kappa\,
k(z)\del_\nu\wh F^{z\nu}
+\frac{N_c}{128\pi^2}\epsilon^{z \mu_2\cdots \mu_5}
\left(
F^a_{\mu_2\mu_3}F^a_{\mu_4\mu_5}
+\wh F_{\mu_2\mu_3}\wh F_{\mu_4\mu_5}
\right)
=0\ ,
\label{eom3}\\
&-\kappa\, k(z)
\left(D_\nu F^{z\nu}\right)^a
+\frac{N_c}{64\pi^2}\epsilon^{z \mu_2\cdots \mu_5}
F^a_{\mu_2\mu_3}\wh F_{\mu_4\mu_5}=0\ ,
\label{eom4}
\end{align}
where $D_\alpha=\del_\alpha+iA_\alpha$ is the covariant derivative.
The baryon in this model corresponds to a soliton
with a nontrivial instanton number on the four-dimensional
space parameterized by $x^M$ ($M=1,2,3,z$).
The instanton number is interpreted as the baryon number
$N_B$, where
\begin{eqnarray}
 N_B=\frac{1}{64\pi^2}\int
d^3x dz\, \epsilon_{M_1M_2M_3M_4} F^a_{M_1M_2}F^a_{M_3M_4}
\ .
\label{NB}
\end{eqnarray}

Unfortunately, because the equations of motion are complicated
nonlinear differential equations in a curved space-time,
it is difficult to find an analytic solution corresponding
to the baryons. However,
as observed in Refs.~\citen{HSSY} and \citen{Hong-Rho-Yee-Yi},
the center of the instanton solution is
located at $z=0$ and its size is of order
$\lambda^{-1/2}$, and hence we can focus on a
tiny region around $z=0$ for large $\lambda$,
in which the warp factors $h(z)$ and $k(z)$
can be approximated by $1$.
Then it follows that the static baryon configuration is given
by the BPST instanton solution with the $U(1)$ electric
field of the form
\begin{align}
A_M^{\rm cl}=&-if(\xi)g\del_M g^{-1} \ ,~~
\wh A_0^{\rm cl}=
\frac{N_c}{8\pi^2 \kappa}
\frac{1}{\xi^2}
\left[1-\frac{\rho^4}{(\rho^2+\xi^2)^2}
\right]\ ,~~~
A_0=\wh A_M=0 \ .
\label{HSSYsol}
\end{align}
Here
\begin{align}
f(\xi)=\frac{\xi^2}{\xi^2+\rho^2}\ ,~~~
g(x)=\frac{(z-Z)-i(\vec{x}-\vec{X})\cdot\vec\tau}{\xi} \ ,~~
\xi=
\sqrt{(z-Z)^2+|\vec{x}-\vec{X}|^2} \ ,
\label{defg}
\end{align}
with $X^M=(X^1,X^2,X^3,Z)=(\vec X,Z)$ being the position
of the soliton in the spatial $\bR^4$ direction
and the instanton size $\rho$.
Substituting this configuration into the action and
taking the nontrivial $z$ dependence of the background
into account as a $1/\lambda$ correction, we find that
$\rho$ and $Z$ have a potential of the form
\begin{eqnarray}
 U(\rho,Z)=8\pi^2\kappa
\left(
1+\frac{\rho^2}{6}
+\frac{N_c^2}{5(8\pi^2\kappa)^2}\frac{1}{\rho^2}
+\frac{Z^2}{3}
\right) \ ,
\label{urhoz}
\end{eqnarray}
which is minimized at
\begin{eqnarray}
\rho_{\rm cl}^2
=\frac{N_c}{8\pi^2\kappa}\sqrt{\frac{6}{5}}\ ,
~~~
Z_{\rm cl}=0\ .
\label{rcl1}
\end{eqnarray}

In order to quantize the soliton, we 
use the moduli space approximation \linebreak method \cite{GeSa,Manton}
and then the system is reduced to the quantum mechanics
on the instanton moduli space.
The $SU(2)$ one-instanton moduli space
is simply given by $\cM\simeq \bR^4\times \bR^4/\bZ_2$,
which is parameterized by $(\vec X,Z)$
and $y_I$ ($I=1,2,3,4$) with the $\bZ_2$ action
$y_I\ra -y_I$. The size of the instanton $\rho$
is related to $y_I$ by $\rho=\sqrt{y_1^2+\cdots+y_4^2}$,
and $a_I\equiv y_I/\rho$ represent the
$SU(2)$ orientations of the instanton.
The Lagrangian of the collective motion
of the soliton was obtained in Ref.~\citen{HSSY} as
\begin{eqnarray}
 L=\frac{M_0}{2}(\dot{\vec{X}}^2+\dot Z^2)+
M_0\, \dot y_I^2-U(\rho,Z)\ ,
\end{eqnarray}
where $M_0=8\pi^2\kappa$.
The Hamiltonian is given by
\begin{eqnarray}
 H=\frac{1}{2M_0}(\vec P^2+P_Z^2)
+\frac{1}{4M_0}\Pi_I^2+U(\rho,Z)
\label{Ham}
\end{eqnarray}
with the canonical momenta
\begin{eqnarray}
\vec P=M_0\dot{\vec X}=-i\frac{\del}{\del\vec X}\ ,~~
P_Z=M_0\dot{Z}=-i\frac{\del}{\del Z}\ ,~~
\Pi_I=2M_0\dot{y_I}=-i\frac{\del}{\del y_I}\ .
\label{mom}
\end{eqnarray}

This system is manifestly
invariant under $SO(4)$ rotation acting on $y^I$.
As argued in Refs.~\citen{ANW} and \citen{HSSY} the generators of
$SO(4)\simeq (SU(2)_I\times SU(2)_J)/\bZ_2$ symmetry
correspond to the isospin and spin operators
given by
\begin{eqnarray}
I_a=\frac{i}{2}\left(
y_4\frac{\del}{\del y_a}-y_a\frac{\del}{\del y_4}
-\epsilon_{abc}\,y_b\frac{\del}{\del y_c}
\right)\ ,~~
J_a=\frac{i}{2}\left(
-y_4\frac{\del}{\del y_a}+y_a\frac{\del}{\del y_4}
-\epsilon_{abc}\,y_b\frac{\del}{\del y_c}
\right)\ ,
\nn\\
\label{IJ}
\end{eqnarray}
respectively.

The explicit
eigenfunctions of the Hamiltonian (\ref{Ham})
are obtained in Ref.~\citen{HSSY}. They are characterized by
the quantum numbers $(l,I_3,s,n_\rho,n_z)$ as well as
the momentum $\vec p$. Here $l=1,3,5,\cdots$ are
positive odd integers related to isospin $I$ and
spin $J$ by $I=J=l/2$. Note that (\ref{IJ}) implies
$\vec I^2=\vec J^2$, and hence only the states
with $I=J$ appear in the spectrum as in Ref.~\citen{ANW}
for the Skyrme model.
$I_3$ and $s$ denote the eigenvalues of
the third component of the isospin
and spin, respectively. $n_\rho$ and $n_z$ are nonnegative
integers corresponding to the excitations with respect
to $\rho$ and $Z$, respectively. The species of the baryon
are specified by $B\equiv (l,I_3,n_\rho,n_z)$ and
the wavefunctions are of the form
\begin{eqnarray}
 \ket{\vec p,B,s}= \ket{\vec p}\ket{B,s}\ ,
\end{eqnarray}
with $\ket{\vec p}=\frac{1}{(2\pi)^{3/2}}e^{i\vec p\cdot\vec X}$.
For later use, we normalize the baryon state as
\begin{eqnarray}
 \vev{B,s|B',s'}=\delta_{B,B'}\delta_{s,s'}\ .
\label{normalize}
\end{eqnarray}
For example, the proton and neutron
are interpreted as the particles with
$B=(1,1/2,0,0)$ and $B=(1,-1/2,0,0)$, respectively,
and the corresponding wavefunctions with $s=1/2$ are
\begin{eqnarray}
 \ket{\,p\uparrow\,}\propto
R(\rho)\psi_Z(Z)(a_1+ia_2)\ ,~~~
 \ket{\,n\uparrow\,}\propto
R(\rho)\psi_Z(Z)(a_4+ia_3)\ ,
\label{wf}
\end{eqnarray}
respectively, where
\begin{eqnarray}
 R(\rho)=\rho^{-1+2\sqrt{1+N_c^2/5}}
e^{-\frac{M_0}{\sqrt{6}}\rho^2}\ ,~~
\psi_Z(Z)=e^{-\frac{M_0}{\sqrt{6}}Z^2}\ ,
\end{eqnarray}
up to normalization constants.
See Appendix \ref{ap:wv} for more details.

The excitation numbers $(n_\rho,n_z)$ are the quantum numbers 
which Skyrmions cannot have, and are thus peculiar to
the instanton picture of baryons obtained in Ref.~\citen{HSSY}. 
For example, $B=(1,\pm 1/2, 1,0)$ corresponds to the Roper excitation
$N(1440)$, and $B=(1,\pm 1/2, 0,1)$ corresponds to $N(1535)$. 
Higher spin baryons are also included, such as
 $B=(3,I_3,0,0)$ with $I_3=\pm 3/2,\pm 1/2$ 
giving $\Delta(1232)$.

\subsection{Currents}

As in the usual effective field theory approach to the hadrons,
it is useful to introduce the external gauge fields
$\cA_{L\mu}$ and $\cA_{R\mu}$ associated with the
chiral symmetry  $U(N_f)_L\times U(N_f)_R$.
The currents $\cJ_L^\mu$ and $\cJ_R^\mu$ associated
with the chiral symmetry are then read off from 
the terms linear
with respect to the external gauge fields
in the effective action as
\begin{align}
S\,\big|_{\cO(\cA_L,\,\cA_R)}=-2\int d^4x\tr\Big(
\cA_{L\mu}\cJ_{L}^\mu+\cA_{R\mu}\cJ_{R}^\mu
\Big) \ .
\label{Scurr}
\end{align}
In terms of the vector and axial-vector current
\begin{align}
 \cJ_{V}^\mu=\cJ_L^\mu+\cJ_R^\mu \ ,~~~
 \cJ_{A}^\mu=\cJ_L^\mu-\cJ_R^\mu \ ,
\end{align}
(\ref{Scurr}) becomes
\begin{align}
S\,\big|_{\cO(\cA_L,\,\cA_R)}
=-2\int d^4x\tr\Big(\cV^{(+)}_\mu\cJ_{V}^\mu
+\cV_\mu^{(-)}\cJ_{A}^\mu 
\Big) \ ,
\label{VA}
\end{align}
where
\begin{align}
 \cV_\mu^{(\pm)}=
\frac{1}{2}\left(\cA_{L\mu}\pm\cA_{R\mu}\right)
\end{align}
are external vector and axial-vector fields.

We can apply this idea to compute the currents in 
the present model,
which leads us to a prescription analogous to that used
in the AdS/CFT correspondence \cite{Gubser:1998bc,Witten:1998qj},
as first explored in Ref.~\citen{Son-Stephanov}
and discussed recently in Ref.~\citen{Hata-Murata-Yamato}.
As studied in Refs.~\citen{SaSu1} and \citen{SaSu2}, the chiral symmetry 
is identified with the constant gauge transformation
at infinity $z=\pm\infty$, and
the external gauge fields $\cA_{L\mu}$ and $\cA_{R\mu}$
can be introduced by considering the five-dimensional
gauge field with the boundary conditions
\begin{align}
 \cA_\mu(x^\mu,z\to +\infty)=\cA_{L\mu}(x^\mu) \ ,~~
 \cA_\mu(x^\mu,z\to -\infty)=\cA_{R\mu}(x^\mu) \ ,
\label{bdry}
\end{align}
respectively.
To compute the currents,
we substitute a solution of the equations of motion
with this boundary condition into the action
and keep only the linear terms in the external
gauge fields $\cA_{L}$ and $\cA_R$
assuming that the external fields are infinitesimal.
Consider the gauge configuration of the form
\begin{eqnarray}
&&\cA_{\alpha}(x^\mu,z)= 
\cA^{\rm cl}_{\alpha}(x^\mu,z)+\delta \cA_{\alpha}(x^\mu,z)\ ,
\label{A}
\end{eqnarray}
where $\cA^{\rm cl}_\alpha$ is a classical solution of the
equations of motion with $\cA_\alpha^{\rm cl}(z=\pm\infty)=0$
and $\delta \cA_\alpha(x,z)$ is an infinitesimal
deviation from it satisfying
\begin{eqnarray}
\delta \cA_\mu(x^\mu,z\to +\infty)=\cA_{L\mu}(x^\mu) \ ,~~
\delta \cA_\mu(x^\mu,z\to -\infty)=\cA_{R\mu}(x^\mu) \ .
\label{dA}
\end{eqnarray}
Substituting this configuration into the action
(\ref{model}), we obtain
\begin{align}
S\,\big|_{\cO(\cA_L,\,\cA_R)}=
\kappa\int d^4x\,2\tr\Big[
\tr\left(\delta\cA^{\mu}k(z)\cF^{\rm cl}_{\mu z}\right)
\Big]^{z=+\infty}_{z=-\infty} \ ,
\label{surface}
\end{align}
which implies
\begin{align}
&\cJ_{L\mu}=
-\kappa
\left(k(z)\,\cF^{\rm cl}_{\mu z}\right)\Big |_{z=+\infty}\ ,~~
\cJ_{R\mu}=
+\kappa
\left(k(z)\,\cF^{\rm cl}_{\mu z}\right)\Big |_{z=-\infty}\ ,
\label{def:current}
\\
&\cJ_{V\mu}=-\kappa 
\Big[\,k(z)\cF^{\rm cl}_{\mu z}\Big]^{z=+\infty}_{z=-\infty}\ ,~~
\cJ_{A\mu}=-\kappa 
\Big[\,\psi_0(z)k(z)\cF^{\rm cl}_{\mu z}\Big]^{z=+\infty}_{z=-\infty}
\ ,
\label{current2}
\end{align}
where $\psi_0(z)=\frac{2}{\pi}\arctan z$.
Here only the surface terms at $z=\pm\infty$
remain in (\ref{surface})
because of the equations of motion.

Note that the currents in (\ref{def:current}) are consistent with
the four-dimensional effective action obtained
in Ref.~\citen{SaSu2}. It was shown in
Ref.~\citen{SaSu2}\footnote{See Eq.~(5.12) in Ref.~\citen{SaSu2}.}
that the four-dimensional effective action derived from
our model has the following terms:
\begin{eqnarray}
S\,\big|_{\cO(\cA_L,\,\cA_R)}=\int d^4 x\,
2\tr\left(
\cV^{(+)\mu} \sum_{n=1}^\infty
g_{v^n}v_\mu^n 
+\cV^{(-)\mu} \left(\sum_{n=1}^\infty
g_{a^n}a_\mu^n 
+f_\pi\del_\mu\Pi \right)
\right)\ ,
\label{4dim}
\end{eqnarray}
where $\Pi(x)$, $v_\mu^{n}(x)$, and $a_\mu^{n}(x)$
are the pion, vector meson, and axial-vector meson fields, respectively,
and $f_\pi$, $g_{v^n}$, and $g_{a^n}$ are the decay constants
of these mesons, respectively.
These mesons are related to the five-dimensional gauge field as
\begin{eqnarray}
&&\cA_\mu(x,z)=
\sum_{n=1}^\infty v_\mu^{n}(x)\psi_{2n-1}(z)
+\sum_{n=1}^\infty a_\mu^{n}(x)\psi_{2n}(z)\ ,~~~
\cA_z(x,z)=\Pi(x)\phi_0(z)\ ,
\label{exp}
\end{eqnarray}
where $\phi_0=\frac{1}{\sqrt{\kappa\pi}}\frac{1}{k(z)}$
and $\{\psi_n(z)\}_{n=1,2,\cdots}$
is a complete set of the functions of $z$
consisting of the eigenfunctions of the eigenequation
\begin{eqnarray}
-h(z)^{-1}\del_z\big(k(z)\del_z\psi_n\big)=\lambda_n\psi_n\ ,
\label{psi1}
\end{eqnarray}
with the normalization condition
\begin{eqnarray}
\kappa\int dz\,h(z)\psi_n\psi_m=\delta_{mn}\ .
\label{psi2}
\end{eqnarray}
We can show that $\psi_{2n-1}(z)$ and $\psi_{2n}(z)$
are even and odd functions of $z$, respectively,
if we arrange $\psi_n(z)$ such that the eigenvalues
satisfy $\lambda_1<\lambda_2<\lambda_3<\cdots$.
The decay constants are given by
\begin{eqnarray} 
f_\pi=2\sqrt{\frac{\kappa}{\pi}}\ ,~~
g_{v^n}=-2\kappa\left(
k(z)\del_z\psi_{2n-1}
\right)\big|_{z=+\infty}\ ,~~
g_{a^n}=-2\kappa\left(
k(z)\del_z\psi_{2n}
\right)\big|_{z=+\infty}\ .
\nn\\
\label{const}
\end{eqnarray}
Note that the functions
$\psi_n(z)$ behave as $\cO(z^{-1})$ 
in the $z\ra\pm\infty$ limit and
the decay constants $g_{v^n}$ and $g_{a^n}$
are determined as the coefficients in front of $1/z$
as
\begin{align}
\psi_{2n-1}(z\to\pm\infty)\simeq
\pm\frac{g_{v^n}}{2\kappa z}
\ ,~~
\psi_{2n}(z\to\pm\infty)\simeq
\frac{g_{a^n}}{2\kappa z} \ .
\end{align}
The following expressions are also useful:
\begin{eqnarray}
g_{v^n}=\lambda_{2n-1}\kappa
\int dz\,h(z)\psi_{2n-1}\ ,~~
g_{a^n}=\lambda_{2n}\kappa
\int dz\,h(z)\psi_{2n}\psi_0\ ,
\label{gvga2}
\end{eqnarray}
which are obtained by using (\ref{const}) and
(\ref{psi1}).

Comparing (\ref{VA}) and (\ref{4dim}),
the vector and axial-vector currents are obtained as
\begin{eqnarray}
&&\cJ_{V\mu}
=-\sum_{n=1}^\infty
g_{v^n}v_\mu^{n}\ ,~~~
\cJ_{A\mu}
=-f_\pi\del_\mu\Pi
-\sum_{n=1}^\infty
g_{a^n}a_\mu^{n}\ .
\label{4dimcurrent}
\end{eqnarray}
Using (\ref{exp}) and (\ref{const}),
we can easily check that (\ref{current2})
and (\ref{4dimcurrent}) are equivalent.
Note that the vector current is expressed by the
vector mesons $v_\mu^n$, 
which is a direct consequence of the complete vector
meson dominance of the model found in Ref.~\citen{SaSu2}.
(See also Refs.~\citen{Son-Stephanov} and \citen{HoYoSt}.)

As in (\ref{decom}), we decompose the 
currents as
\begin{eqnarray}
\cJ_\mu
=J_{\mu}+\wh J_{\mu}\frac{{\bf 1}_2}{2}
=J^a_{\mu}\frac{\tau^a}{2}
+\wh J_{\mu}\frac{{\bf 1}_2}{2}
=\sum_{C=0}^3\cJ_\mu^C\,\frac{\tau^C}{2} \ .
\end{eqnarray}
Then the baryon number current is given by
\begin{eqnarray}
 J_B^\mu=\frac{2}{N_c} \wh J_V^\mu
=-\frac{2}{N_c}\kappa 
\left[k(z)\wh F^{\mu z}\right]^{z=+\infty}_{z=-\infty}
\ .
\label{JB}
\end{eqnarray}
As a check, the baryon number density
is computed as
\begin{eqnarray}
J_B^0
&=&-\frac{2}{N_c}\kappa \int dz\,
\del_z(k(z)\wh F^{0z})\nn\\
&=&-\frac{1}{64\pi^2}\int dz\,
\epsilon^{0\,M_1M_2M_3M_4} F^a_{M_1M_2}F^a_{M_3M_4}
+(\mbox{total derivative})\ ,
\end{eqnarray}
where we have used the equation of motion (\ref{eom1}).
This is consistent with (\ref{NB})
as expected.

\subsection{Asymptotic solution}

In order to calculate the currents (\ref{def:current}),
we have to know how the field strength $\cF_{\mu z}$ behaves
at $z=\pm\infty$. Unfortunately, we cannot directly use
the solution (\ref{HSSYsol}) since it is only valid
in the region $\xi\ll 1$.\footnote{
Here we are assuming $Z\sim Z_{\rm cl}=0$,
since the expectation value of $f(Z)$ is approximated by its classical
value $f(Z_{\rm cl})$ for the large $N_c$ and large $\lambda$ limit.}
In this subsection,
we study how to extend the solution to the $1\ll\xi$ region.

Let us first summarize the gauge configuration
in the $\xi\ll 1$ region.
Here we include the time-dependent moduli parameters,
which are treated as operators in quantum mechanics
(\ref{Ham}).
As explained in Ref.~\citen{HSSY}, the $SU(2)$ gauge field
takes the form
\begin{eqnarray}
A_M=VA_M^{\rm cl}V^{-1}-iV\del_M V^{-1} \ ,
\label{HSSY:AM}
\end{eqnarray}
where $A_M^{\rm cl}$ is the solution (\ref{HSSYsol})
and $V$ satisfies
\begin{eqnarray}
-iV^{-1}\dot V
=-\dot X^M A_M^{\rm cl}+\chi^a\,f(\xi)\,
g\frac{\tau^a}{2}g^{-1}\ ,
\end{eqnarray}
which follows from the Gauss law constraint.
Here
\begin{eqnarray}
\chi^a=
-i\tr\left(\tau^a\ba^{-1}\dot\ba\right)
=-\frac{i}{\rho^2}
\tr\left(\tau^a \by^\dag \dot\by\right)
=\frac{1}{4\pi^2\kappa\rho^2}J_a
\ ,
\label{chi}
\end{eqnarray}
where $\by=y_4+iy_a\tau^a$, $\ba=a_4+ia_a\tau^a=\by/\rho$,
and $J_a$ is the spin operator defined in (\ref{IJ}).

It is convenient to perform the gauge transformation
\begin{eqnarray}
A_\alpha&\ra& A^G_\alpha\equiv
GA_\alpha G^{-1}-iG\del_\alpha G^{-1}\ ,
\end{eqnarray}
with $G=\ba\, g^{-1}V^{-1}$. Then,
\begin{align}
A_0^G&=-i(1-f(\xi))\,\ba\,\dot\ba^{-1}
+i(1-f(\xi))\,\dot{X}^M\,
\ba\,(g^{-1}\partial_M g)\,\dot\ba^{-1}\ ,
\label{A0G}\\
A_M^G&=
-i(1-f(\xi))\,
\ba\,(g^{-1}\del_M g)\,\ba^{-1}\ .
\label{AMG}
\end{align}
This choice of gauge is useful in considering
the asymptotic behavior,
as we will see in the following,
while a singularity develops at $\xi=0$.

The $U(1)$ part is treated
as a perturbation in the background given by
the $SU(2)$ gauge configuration obtained above.
The leading contribution to the $U(1)$ part
is obtained by solving the following linearized
equations of motion (in the Lorenz gauge):
\begin{align}
\del_M\del^M
\wh A^0&=
\frac{3}{\pi^2a\lambda}\frac{\rho^4}{(\xi^2+\rho^2)^4}\ ,
\label{lin1-2}\\
\del_M\del^M\wh A^i
&=
\frac{3}{\pi^2a\lambda}\frac{\rho^4}{(\xi^2+\rho^2)^4}
\left(\dot X^i+\frac{\chi^a}{2}(\epsilon^{iaj}x^j-\delta^{ia}z)
+\frac{\dot\rho\, x^i}{\rho}\right)\ ,
\label{lin2-2}\\
\del_M\del^M\wh A_z
&=
\frac{3}{\pi^2a\lambda}
\frac{\rho^4}{(\xi^2+\rho^2)^4}
\left(\dot Z+\frac{\chi^ax^a}{2}
+\frac{\dot\rho\, z}{\rho}\right)\ .
\label{lin3-2}
\end{align}
Here we substituted (\ref{A0G}) and (\ref{AMG})
into the equations of motion (\ref{eom1}) and
(\ref{eom3})
with the warp factors $h(z)$ and $k(z)$ approximated
by 1.
We neglect the terms including $\partial_0^2$,
because we are interested in slowly
moving solitons.\footnote{
The time derivative squared such as
$\ddot{Z}$ and $\dot{Z}^2$ can be traded by
$Z$ or $Z^2$  via its Schr\"odinger equation or
the relation $\dot Z= P_Z/M_0$; thus
for large $N_c$ and $\lambda$,
it is approximated by its classical value, which vanishes.}

The regular solution is found to be 
\begin{align}
\wh A_0&=
\frac{1}{8\pi^2 a\lambda}\frac{1}{\xi^2}
\left[1-\frac{\rho^4}{(\xi^2+\rho^2)^2}
\right]
=
\frac{1}{8\pi^2 a\lambda}
\frac{\xi^2+2\rho^2}{(\xi^2+\rho^2)^2} \ ,
\label{whA0}\\
\wh A_i&=
-\frac{1}{8\pi^2 a\lambda}
\left[
\frac{\xi^2+2\rho^2}{(\xi^2+\rho^2)^2}
\dot X^i
+\frac{\rho^2}{(\xi^2+\rho^2)^2}
\left(\frac{\chi^a}{2}(\epsilon^{iaj}x^j-\delta^{ia}z)
+\frac{\dot\rho\, x^i}{\rho}
\right)
\right]\ ,
\label{whAi}\\
\wh A_z&=
-\frac{1}{8\pi^2 a\lambda}
\left[
\frac{\xi^2+2\rho^2}{(\xi^2+\rho^2)^2}
\dot Z
+\frac{\rho^2}{(\xi^2+\rho^2)^2}
\left(\frac{\chi^ax^a}{2}
+\frac{\dot\rho\, z}{\rho}
\right)
\right]\ .
\label{whAz}
\end{align}
Note that (\ref{whAi}) and (\ref{whAz})
are neglected in Ref.~\citen{HSSY},
since the energy contributions from them
are subleading in the
$1/\lambda$ expansion given in Ref.~\citen{HSSY}.
However, here we keep them because
they give the leading contribution
to the isoscalar current density.
It can also be shown that
the $\wh F^2$ terms in the equations
of motion (\ref{eom1}) and (\ref{eom3})
are subleading in the $1/\lambda$ expansion,
justifying the above perturbative
treatment for the $U(1)$ part.

So far, we have established
a solution that is valid in the region $\xi\ll 1$.
We now consider how to find the solution in the
$1\ll \xi$ region. The key observation is that
all the components of the gauge field in
(\ref{A0G}), (\ref{AMG}), (\ref{whA0}),
(\ref{whAi}), and (\ref{whAz})
are suppressed in the $\rho\ll \xi\ll 1$ region
in the large $\lambda$ limit. This implies that
the nonlinear terms in the equations of motion
can be neglected in this region for large $\lambda$.
Our strategy is to find a solution of the
linearized equations of motion in the $\rho\ll \xi$ region
that smoothly connects the previous solution
in the overlapping region $\rho\ll \xi\ll 1$.

For this purpose, we note that for $\rho\ll\xi\ll 1$,
the gauge field
(\ref{A0G}), (\ref{AMG}), (\ref{whA0}), (\ref{whAi}), and (\ref{whAz})
are approximated as
\begin{align}
\wh A_0&\simeq
-\frac{1}{2 a\lambda}G^{\rm flat}(\bx,z;\bX,Z) \ ,
\label{flat1}\\
\wh A_i&\simeq
\frac{1}{2 a\lambda}
\left[
\dot X^i
+\frac{\rho^2}{2}
\left\{\frac{\chi^a}{2}
\left(\epsilon^{iaj}\frac{\del}{\del X^j}
-\delta^{ia}\frac{\del}{\del Z}
\right)
+\frac{\dot\rho}{\rho}
\frac{\del}{\del X^i}
\right\}
\right]
G^{\rm flat}(\bx,z;\bX,Z)
\ ,
\label{flat2}\\
\wh A_z&\simeq
\frac{1}{2 a\lambda}
\left[
\dot Z
+\frac{\rho^2}{2}
\left(\frac{\chi^a}{2}
\frac{\del}{\del X^a}
+\frac{\dot\rho}{\rho}
\frac{\del}{\del Z}
\right)
\right]
G^{\rm flat}(\bx,z;\bX,Z)\ ,
\label{flat3}
\end{align}
\begin{align}
 A_0^G&\simeq4\pi^2\rho^2 i\ba\,\dot\ba^{-1}
\,G^{\rm flat}(\bx,z;\bX,Z)
\nn\\
&~~~+2\pi^2\rho^2
\ba\tau^a\ba^{-1}\left(\dot X^i
\left(\epsilon_{iaj}\frac{\del}{\del X^j}
-\delta^{ai}\frac{\del}{\del Z}
\right)
+\dot Z\frac{\del}{\del X^a}\right)
G^{\rm flat}(\bx,z;\bX,Z)\ 
,\\
A_i^G&\simeq2\pi^2{\rho^2}
\left(\ba\tau^i\ba^{-1}\frac{\del}{\del Z}
+\epsilon_{ija}\ba\tau^a\ba^{-1}\frac{\del}{\del X^j}\right)
G^{\rm flat}(\bx,z;\bX,Z)\ ,
\\
A_z^G&\simeq-2\pi^2{\rho^2}
\ba\tau^a\ba^{-1}\frac{\del}{\del X^a}
G^{\rm flat}(\bx,z;\bX,Z)\ ,
\label{AzG2}
\end{align}
where
\begin{align}
G^{\rm flat}(\bx,z;\bX,Z)=-\frac{1}{4\pi^2}\frac{1}{\xi^2}\ ,
\label{Gflat}
\end{align}
is the Green's function in the flat
$\bR^4$ that satisfies
\begin{align}
\del_M\del^MG^{\rm flat}(\bx,z;\bX,Z)
=\delta^3(\bx-\bX)\delta(z-Z)\ .
\end{align}
We can easily check that the gauge configuration
(\ref{flat1})--(\ref{AzG2}) satisfies
the Maxwell and the linearized YM equations
without sources:
\begin{align}
 \partial_\beta\wh F^{\alpha\beta}=
 \partial_\beta F^{\alpha\beta}\big|_{\rm linear}=0 \ ,
\label{linear}
\end{align}
as well as the gauge condition
\begin{align}
\del^\alpha \wh A_\alpha=0 \ ,~~
\del^\alpha A_\alpha^G=0 \ .
\label{gauge}
\end{align}

In order to connect this solution to the
large $\xi$ region,
we have to take into account the effect of the
curved background.
The linearized equations of motion and the gauge condition
that generalize (\ref{linear}) and (\ref{gauge})
to the case with nontrivial warp factors $h(z)$
and $k(z)$ are
\begin{align}
&h(z)\del_\mu^2\wh A_i+\del_z (k(z)\del_z\wh A_i)=0\ ,~~~
\del_\mu^2\wh A_z+\del_z (h(z)^{-1}\del_z (k(z)\wh A_z))=0\ ,
\label{cur1}\\
&h(z)\del_\mu^2A_i^G+\del_z (k(z)\del_zA_i^G)=0\ ,~~~
\del_\mu^2A_z^G+\del_z (h(z)^{-1}\del_z (k(z)A_z^G))=0\ ,
\label{cur2}
\end{align}
and
\begin{eqnarray}
h(z)\del^\mu \wh A_\mu+\del_z (k(z)\wh A_z)=0\ ,~~~
h(z)\del^\mu A^G_\mu+\del_z (k(z)A_z^G)=0\ ,
\label{cur3}
\end{eqnarray}
respectively.

To solve these equations, we
define Green's functions in the curved space as
\begin{align}
 G(\bx,z;\bX,Z)&=\kappa\sum_{n=1}^\infty
\psi_n(z)\psi_n(Z)Y_n(|\bx-\bX|)\ ,
\label{G1}\\
H(\bx,z;\bX,Z)&=\kappa\sum_{n=0}^\infty
\phi_n(z)\phi_n(Z)Y_n(|\bx-\bX|)\ ,
\label{H1}
\end{align}
where
 $\{\psi_n(z)\}_{n=1,2,\cdots}$ is the complete set
defined in (\ref{psi1}) and (\ref{psi2}),
$\{\phi_n(z)\}_{n=0,1,\cdots}$ is another complete set
given by
\begin{eqnarray}
\phi_0(z)=\frac{1}{\sqrt{\kappa\pi}}\frac{1}{k(z)}\ ,~~~
\phi_n(z)=\frac{1}{\sqrt{\lambda_n}}\del_z\psi_n(z)\ ,
~~~(n=1,2,\cdots)
\label{phi1}
\end{eqnarray}
and $Y_n(r)$ is the Yukawa potential
with meson mass $m_n=\sqrt{\lambda_n}$,
\begin{align}
 Y_n(r)=-\frac{1}{4\pi}\frac{e^{-\sqrt{\lambda_n}\,r}}{r}
\ ,
\label{Yukawa}
\end{align}
which satisfies
\begin{eqnarray}
(\del_i^2-\lambda_n)Y_n(|\bx-\bX|)=
\delta^3(\bx-\bX)\ .
\label{Yn}
\end{eqnarray}
Note that the normalization of the functions
 $\{\phi_n\}_{n=0,1,2,\cdots}$ is fixed by
\begin{eqnarray}
\kappa\int dz\, k(z)\phi_n\phi_m=\delta_{mn}\ .
\end{eqnarray}
Using (\ref{psi1}), (\ref{phi1}), (\ref{Yn}), and
the completeness conditions
\begin{eqnarray}
 \kappa\, h(z)\sum_{n=1}^\infty
\psi_n(z)\psi_n(Z)=\delta(z-Z)\ ,~~
 \kappa\, k(z)\sum_{n=1}^\infty
\phi_n(z)\phi_n(Z)=\delta(z-Z)\ ,
\label{complete}
\end{eqnarray}
it is easy to verify
\begin{align}
h(z)\del_i^2G+\del_z (k(z) \del_zG)=&\,
\delta^3(\bx-\bX)\delta(z-Z)\ ,
\label{G}\\
\del_i^2 H+\del_z (h(z)^{-1}\del_z (k(z)H))
=&\,k(z)^{-1}\,
\delta^3(\bx-\bX)\delta(z-Z)\ ,
\label{H}\\
\del_z (k(z)H)+h(z)\del_ZG
=&\,0\ .
\label{HG1}
\end{align}

Then, the solution of equations
(\ref{cur1})--(\ref{cur3}) is obtained by
replacing the Green's function $G^{\rm flat}$ in 
(\ref{flat1})--(\ref{AzG2})
with $G$ or $H$ as follows:
\begin{align}
\wh A_0&\simeq
-\frac{1}{2 a\lambda}G(\bx,z;\bX,Z)\ ,
\label{curved1}
\\
\wh A_i&\simeq
\frac{1}{2 a\lambda}
\left[
\dot X^i
+\frac{\rho^2}{2}
\left\{\frac{\chi^a}{2}
\left(\epsilon^{iaj}\frac{\del}{\del X^j}
-\delta^{ia}\frac{\del}{\del Z}
\right)
+\frac{\dot\rho}{\rho}
\frac{\del}{\del X^i}
\right\}
\right]
G(\bx,z;\bX,Z)
\ ,
\label{curved2}
\\
\wh A_z&\simeq
\frac{1}{2 a\lambda}
\left[\dot Z
+\frac{\rho^2}{2}
\left(
\frac{\chi^a}{2}\frac{\del}{\del X^a}
+\frac{\dot\rho}{\rho}\frac{\del}{\del Z}
\right)
\right]
H(\bx,z;\bX,Z)\ ,
\label{curved3}
\end{align}
\begin{align}
A_0^G&\simeq4\pi^2\rho^2 i\ba\,\dot\ba^{-1}
\,G(\bx,z;\bX,Z)\nn\\
&~~~+2\pi^2\rho^2
\ba\tau^a\ba^{-1}\left(\dot X^i
\left(\epsilon_{iaj}\frac{\del}{\del X^j}
-\delta^{ai}\frac{\del}{\del Z}
\right)
+\dot Z\frac{\del}{\del X^a}\right)
G(\bx,z;\bX,Z)\ ,
\label{A0Gc}\\
A_i^G&\simeq-2\pi^2{\rho^2}
\ba\tau^a\ba^{-1}\left(
\epsilon^{iaj}\frac{\del}{\del X^j}-\delta^{ia}\frac{\del}{\del Z}\right)
G(\bx,z;\bX,Z)\ ,
\label{AiGc}\\
A_z^G&\simeq-2\pi^2{\rho^2}
\ba\tau^a\ba^{-1}\frac{\del}{\del X^a}
H(\bx,z;\bX,Z)\ .
\label{AzGc}
\end{align}
Here,
we neglected the terms including $\partial_0^2$
as before.

Since the Green's functions $G$ and $H$ approach
$G^{\rm flat}$ 
for  $\rho\ll\xi\ll 1$,
this solution 
is smoothly connected with
the previous solution (\ref{flat1})--(\ref{AzG2})
in this region, as expected.
Note also that the Green's functions $G$ and $H$ vanish
at $\xi\ra\infty$, and hence the linear approximation
of the equations of motion does not break down all the way
to infinity.
Therefore, we can read off
the behavior of the gauge field at $z\ra\pm\infty$,
which is needed to calculate the currents, from 
(\ref{curved1})--(\ref{AzGc}).

\subsection{Computation of the current}

Since the wavefunctions of the low-lying
baryon states, such as (\ref{wf}),
are dominant for
 $Z\sim \cO(\lambda^{-1/2}N_c^{-1/2})\ll 1$,
we can use the approximation
$h(Z)\simeq k(Z)\simeq 1$ and the relation
\begin{eqnarray}
-\del_Z^2\psi_n(Z)\simeq \lambda_n\psi_n(Z)\ ,
\end{eqnarray}
which follows from (\ref{psi1}). This implies
\begin{eqnarray}
\del_Z H+\del_z G\simeq 0\ ,~~
(\del_i^2+\del_Z^2) G\simeq 0\ ,~~
(\del_i^2+\del_Z^2) H\simeq 0\ .
\label{HG}
\end{eqnarray}

Using the asymptotic
solution (\ref{curved1})--(\ref{AzGc})
and the relations (\ref{HG}),
we obtain
\begin{align}
\wh F_{0z}&\simeq \frac{1}{2a\lambda}\del_zG \ ,
\\
\wh F_{iz}&\simeq \frac{1}{2a\lambda}
\Big[\dot Z\del_i H-\dot X^i\del_z G
-\frac{\rho^2\chi^a}{4}\left(
(\del_i\del_a-\delta^{ia}\del_j^2) H
-\epsilon^{iaj}\del_j\del_z G
\right)\Big] \ , 
\\
F_{0z}&\simeq
2\pi^2\del_0(\rho^2\ba\tau^a\ba^{-1})\del_aH
-4\pi^2\rho^2i\ba\,\dot\ba^{-1}\del_zG
\nn\\
&~~~~~
-2\pi^2\rho^2\ba\tau^a\ba^{-1}\dot X^i\left(
(\del_a\del_i-\delta^{ia}\del_j^2) H-
\epsilon^{iaj}\del_j\del_z G
\right) \ ,
\\
F_{iz}&\simeq
2\pi^2\rho^2\ba\tau^a\ba^{-1}
\left((\del_i\del_a-\delta^{ia}\del_j^2) H-
\epsilon^{iaj}\del_j\del_zG
\right)
\end{align}
for $Z\ll 1\ll z$.

The following formulas are also useful:
\begin{align}
G^V(Z,r)&\equiv 
\Big[k(z)\del_z G\Big]^{z=+\infty}_{z=-\infty}
=-\sum_{n=1}^\infty g_{v^n}
\psi_{2n-1}(Z)Y_{2n-1}(r)
\ ,\\
G^A(Z,r)&\equiv 
\Big[\psi_0(z)k(z)\del_z G\Big]^{z=+\infty}_{z=-\infty}
=-\sum_{n=1}^\infty g_{a^n}
\psi_{2n}(Z)Y_{2n}(r)
\ ,\\
H^V(Z,r)
&\equiv
\Big[k(z)H\Big]^{z=+\infty}_{z=-\infty}
=-\sum_{n=1}^\infty \frac{g_{v^n}}{\lambda_{2n-1}}
\del_Z\psi_{2n-1}(Z)Y_{2n-1}(r)
\ ,\\
H^A(Z,r)&\equiv 
\Big[\psi_0(z)k(z)H\Big]^{z=+\infty}_{z=-\infty}
=-\frac{1}{2\pi^2}\frac{1}{k(Z)}\frac{1}{r}
-\sum_{n=1}^\infty \frac{g_{a^n}}{\lambda_{2n}}
\del_Z\psi_{2n}(Z)Y_{2n}(r)
\ ,
\label{HA}
\end{align}
where $r=|\bx-\bX|$.
Note that $G^V$ and $H^A$ are even functions
with respect to $Z$, while $G^A$ and $H^V$
are odd functions.

Now, we are ready to write down the currents from
(\ref{def:current}) and (\ref{current2}).
The vector and axial-vector currents are obtained as
\begin{align}
\wh J_{V,A}^0&=
\frac{N_c}{2}G^{V,A}\ ,
\label{J0U1}
\\
\wh J_{V,A}^i&=
-\frac{N_c}{2}
\Big[\dot Z\del_i H^{V,A}-\dot X^i G^{V,A}
-\frac{\rho^2\chi^a}{4}\left(
(\del_i\del_a-\delta^{ia}\del_j^2) H^{V,A}
-\epsilon^{iaj}\del_j G^{V,A}
\right)\Big] \ , 
\label{JiU1}
\\
J_{V,A}^0&=
2\pi^2\kappa\,
\Big[\del_0(\rho^2\ba\tau^a\ba^{-1})\del_aH^{V,A}
-2\rho^2i\ba\,\dot\ba^{-1}G^{V,A}
\nn\\
&~~~~~
-\rho^2\ba\tau^a\ba^{-1}\dot X^i\left(
(\del_a\del_i-\delta^{ia}\del_j^2) H^{V,A}-
\epsilon^{iaj}\del_jG^{V,A}
\right)\Big] \ ,
\label{J0}
\\
J_{V,A}^i&=
-2\pi^2\kappa\,\rho^2\ba\tau^a\ba^{-1}
\left((\del_i\del_a-\delta^{ia}\del_j^2) H^{V,A}-
\epsilon^{iaj}\del_j G^{V,A}
\right) \ .
\label{Ji}
\end{align}
Using the relation (\ref{HG}), it is easy to check
that these currents are conserved
(up to terms including $\del_0^2$).\footnote{
Here we neglect the effect of the $U(1)_A$ anomaly,
since it is subleading in the $1/N_c$ expansion.
}

\section{Static properties of baryons}
\label{secstatic}

In this section, we study the static properties of
baryons as applications of the currents
obtained in the previous section.

\subsection{Baryon number density, isoscalar mean square radius}
\label{isosR}

The baryon number density is obtained from
 (\ref{JB}) and (\ref{J0U1}) as
\begin{eqnarray}
J_{B}^0
&=&-\sum_{n=1}^\infty g_{v^n}
\psi_{2n-1}(Z)Y_{2n-1}(r)\ .
\end{eqnarray}
As a check, the baryon number charge is calculated
using this expression as
\begin{align}
N_B
&=\int_0^\infty dr\,4\pi r^2\vev{J^0_B(r)}
\label{intrho}\\
&=
\sum_{n=1}^\infty\frac{g_{v^n}}{\lambda_{2n-1}}
\vev{\psi_{2n-1}(Z)}
\nn\\
&=1\ ,
\label{NBsum}
\end{align}
where $\vev{\cO}=\bra{B,s}\cO\ket{B,s}$
is the expectation value with respect to a baryon
state $\ket{B,s}$. 
Here we have used (\ref{gvga2})
and (\ref{complete}).

The baryon number density per unit $r$ is given by
the integrand of (\ref{intrho}),
\begin{eqnarray}
 \rho_B(r)\equiv 4\pi r^2\vev{J_B^0(r)}
=r\sum_{n=1}^\infty g_{v^n}\,\vev{\psi_{2n-1}(Z)}
\,e^{-\sqrt{\lambda_{2n-1}}\,r}\ .
\end{eqnarray}
Then the isoscalar mean square radius is
\begin{eqnarray}
\vev{r^2}_{I=0}=\int_0^\infty
dr\,r^2\,\rho_B(r)
=6\sum_{n=1}^\infty \frac{g_{v^n}}{\lambda_{2n-1}^2}
\,\vev{\psi_{2n-1}(Z)}\ .
\label{rI0}
\end{eqnarray}

In the large $N_c$ and large $\lambda$ limit,
the baryon wavefunction is localized at $Z=0$,
as we can see from the wavefunction (\ref{wf})
for the nucleon states, and hence the
expectation value $\vev{\psi_{2n-1}(Z)}$
can be approximated by its classical value
$\psi_{2n-1}(Z_{\rm cl})=\psi_{2n-1}(0)$.
In this approximation,
it is possible to evaluate the isoscalar mean
square radius (\ref{rI0}) as follows.
Note that the function
\begin{eqnarray}
F(z)\equiv
6\sum_{n=1}^\infty \frac{g_{v^n}}{\lambda_{2n-1}^2}
\psi_{2n-1}(z)
\end{eqnarray}
satisfies 
\begin{eqnarray}
-\del_z(k(z)\del_z F(z))=6\, h(z)\ ,~~F(z)=F(-z)
\label{F}
\end{eqnarray}
and the boundary condition $F(z)\ra 0$
 $(z\ra\pm\infty)$. The first relation is obtained from
(\ref{psi1}), (\ref{gvga2}), and (\ref{complete}).
The solution of (\ref{F}) is given by
\begin{eqnarray}
F(z)=F_0-\int_0^z dz'\, k(z')^{-1}\int_0^{z'}dz''\,6\, h(z'')\ .
\end{eqnarray}
The constant $F_0$ is fixed by the boundary condition and we obtain
\begin{eqnarray}
F_0=\int_0^\infty
 dz'\, k(z')^{-1}\int_0^{z'}dz''\,6\, h(z'')
\simeq 14.3\ .
\end{eqnarray}
Therefore (\ref{rI0}) can be evaluated as
\begin{eqnarray}
\vev{r^2}_{I=0}=
\vev{F(Z)}\simeq F(0)=F_0\simeq 14.3/\Mkk^2
\label{rI0-2}
\ ,
\end{eqnarray}
and if we use the value (\ref{Mkkkappa}) for $\Mkk$,
we obtain
\begin{eqnarray}
\vev{r^2}_{I=0}^{1/2}
\simeq
 0.785~{\rm fm} \ .
\label{claI0}
\end{eqnarray}
The experimental value is 
$\vev{r^2}_{I=0}^{1/2}|_{\rm exp}\simeq 0.806~{\rm fm}$
\cite{PDG}
and the prediction of the Skyrme model in Ref.~\citen{ANW}
is $0.59~{\rm fm}$.

It is interesting to note that this value
is independent of $\lambda$ and $N_c$.
The $N_c$ independence is consistent with
the analysis of baryons in large $N_c$ QCD \cite{WittenBaryon2}.
The $\lambda$ independence suggests that the
size of the baryon number distribution
is governed by the scale of the vector meson mass
rather than the size of the soliton
 $\rho_{\rm cl}\sim \cO(\lambda^{-1/2})$
for large $\lambda$.

Given the wavefunction of the baryon state,
it is also possible to evaluate the
expectation value $\vev{F(Z)}$ numerically.\footnote{
Note that in the previous section the chiral currents were obtained 
using the approximation $Z\sim 0$, and thus the classical value for $Z$
was used. Here, we simply assume the same chiral current and evaluate it
using the quantum states of the baryons.}
For the nucleon wavefunction given by (\ref{wf}),
we obtain
\begin{eqnarray}
\vev{r^2}_{I=0}^{1/2}
\simeq
 0.742~{\rm fm} \ .
\label{protonismsr}
\end{eqnarray}
This value is the same for the other
states with $n_z=0$, such as $\Delta(1232)$ and $N(1440)$,
since $Z$ dependence of the wavefunction
is the same for these states.
For the states with $n_z=1$, such as $N(1535)$,
we obtain
\begin{eqnarray}
\vev{r^2}_{I=0}^{1/2}
\simeq
 0.699~{\rm fm} \ .
\label{1535ismsr}
\end{eqnarray}
See Appendix \ref{xx} for more details.

\subsection{Isoscalar magnetic moment}
\label{I=0mag}

The isoscalar magnetic moment is defined as
\begin{eqnarray}
\mu^i_{I=0}
=
\frac{1}{2}\epsilon^{ijk}
\int d^3x\,x^j J_B^k
=
\frac{1}{N_c}\epsilon^{ijk}
\int d^3x\,x^j \wh J_V^k\ .
\label{muI0}
\end{eqnarray}
It is easy to see that only the last term in (\ref{JiU1})
contributes to the integral.
Then by evaluating the angular integral in (\ref{muI0}) we obtain
\begin{eqnarray}
\mu_{I=0}^i=-\frac{\rho^2\chi^i}{12}
\int_0^\infty dr\,4\pi r^3\del_r J_B^0(r)\ .
\label{magden}
\end{eqnarray}
Using (\ref{chi}), this can be evaluated as
\begin{eqnarray}
\mu_{I=0}^i=
\frac{\rho^2\chi^i}{4}
=\frac{J^i}{2M_0}\ ,
\label{mu}
\end{eqnarray}
where $J^i$ is the spin operator (\ref{IJ}).
For example, for nucleon states with up spin,
this gives
\begin{eqnarray}
\bra{\,p\!\uparrow\!}\,\mu^i_{I=0}\,\ket{\,p\!\uparrow}
=\bra{\,n\!\uparrow\!}\,\mu^i_{I=0}\,\ket{\,n\!\uparrow}
=\frac{1}{4M_0}\delta^{3i}\ .
\label{muI0-2}
\end{eqnarray}
The $g$ factor is defined as
\begin{eqnarray}
\mu^i=\frac{g}{4M_N}\sigma^i\ ,
\label{gfactor}
\end{eqnarray}
where $\sigma^i$ is the Pauli matrix that acts
on a spin doublet,
and $M_N$ is the nucleon mass.
If we use (\ref{Mkkkappa}) and the experimental
value for the nucleon mass $M_N\simeq 940~{\rm MeV}$,
we have\footnote{
Note that $M_N$ is not calculated in Ref.~\citen{HSSY}
because the total contribution from the zero point energy
of the fluctuations around the soliton solution
is difficult to evaluate.
Here the $g$ factor is computed simply to express
the magnetic moments in the unit of $1/(4M_N)|_{\rm exp}$.
}
\begin{eqnarray}
g_{I=0}=g_p+g_n=M_N/M_0\simeq 1.68\ . 
\label{gI0}
\end{eqnarray}
The experimental value is $g_{I=0}|_{\rm exp}\simeq 1.76$ \cite{PDG}
and the prediction obtained from the Skyrme model in Ref.~\citen{ANW} is
$g_{I=0}|_{\rm ANW}\simeq 1.11$.\footnote{ 
The same expressions as (\ref{muI0-2}) and
(\ref{gI0}) were obtained using the current in
Ref.~\citen{Hata-Murata-Yamato}. However, the numerical values were
different, because the values of $\Mkk$ and $\kappa$ used
in Ref.~\citen{Hata-Murata-Yamato} were different from ours.
}

\subsection{Isovector charge density and charge radii}

The isovector charge density is given by (\ref{J0})
and the isovector charge is evaluated
using (\ref{NBsum}) as
\begin{eqnarray}
Q_V=\int d^3 x\,J_V^0
=-4\pi^2\kappa\rho^2 i\ba\,\dot\ba^{-1}\ .
\label{QV}
\end{eqnarray}
A relation similar to (\ref{chi}),
\begin{eqnarray}
-i\tr\left(\tau^a\ba\,\dot\ba^{-1}\right)
=-\frac{i}{\rho^2}
\tr\left(\tau^a \by\, \dot\by^\dag\right)
=\frac{1}{4\pi^2\kappa\rho^2}I_a
\ ,
\label{iso}
\end{eqnarray}
where $I_a$ is the isospin operator defined in (\ref{IJ}),
implies
\begin{eqnarray}
Q_V^a=I_a\ ,
\end{eqnarray}
as expected.

The isovector charge density per unit $r$
is proportional to the angular integral of
the isovector charge density with the 
normalization condition
$\int_0^\infty dr\,\rho_{I=1}(r)=1$.
This turns out to be
identical to the baryon number density:
\begin{eqnarray}
  \rho_{I=1}(r)=\rho_B(r) \ .
\label{rho01}
\end{eqnarray}
Therefore, the isovector mean square charge radius
$\vev{r^2}_{I=1}=
\int_0^\infty\,r^2\rho_{I=1}(r)dr$ is the same
as the isoscalar mean square radius $\vev{r^2}_{I=0}$
evaluated in \S \ref{isosR}.
This result is somewhat puzzling since it is known
that the isovector mean square radius is
divergent in the chiral limit \cite{BeZe}.
There is however no contradiction.
This divergence is due to the IR divergence
of pion loops \cite{GaSaSv}. Our analysis only
involves a string world-sheet with disk topology, and hence
the pion loops are not included. Therefore it will be
important to include the quark mass in the model\footnote{ 
See Refs.~\citen{Casero:2007ae}--\citen{Dhar:2008um}
for recent developments toward the incorporation of
the quark mass in the model.
}
 and estimate the contribution
from the annulus diagram to make a comparison 
with the experimental value,
which is beyond the scope of the present paper.

The electric charge is defined by
\begin{eqnarray}
 Q_{\rm em}=I_3+\frac{N_B}{2}\ ,
\end{eqnarray}
which gives $Q_{\rm em}=1$ for a proton ($I_3=1/2$)
 and $Q_{em}=0$
for a neutron  ($I_3=-1/2$).
Then, because of identity (\ref{rho01}),
the electric charge density is given
by $\rho_{E}=\rho_{I=0}$
for a proton and  $\rho_{E}=0$ for a neutron,
and we obtain
\begin{eqnarray}
 \vev{r^2}_{E,\rm p}=\vev{r^2}_{I=0}
~~(\mbox{for a proton})\ ,~~~~
 \vev{r^2}_{E,\rm n}=0~~(\mbox{for a neutron})\ .
\label{chr}
\end{eqnarray}
The experimental values in Ref.~\citen{PDG} are
\begin{eqnarray}
 \vev{r^2}_{E,\rm p}\big|_{\rm exp}\simeq (0.875~{\rm fm})^2
\ ,~~~~
 \vev{r^2}_{E,\rm n}\big|_{\rm exp}\simeq-0.116~{\rm fm}^2
\ .
\end{eqnarray}
Although our calculation does not reproduce the experimental value of
the electric charge radius of the neutron
$\vev{r^2}_{E,\rm n}$, the vanishing of the neutron electric charge density
seems to be a good approximation for reproducing the observed
behavior of the electric form factor, as we will study
in \S 4.

The charge radius for the excited baryons is similarly found 
to be the same as their
isoscalar mean square radius $\vev{r^2}_{I=0}$ obtained 
in \S \ref{isosR}. In particular, our analysis predicts that the
Roper excitation $N(1440)$ has a charge radius equal to that of a
proton (\ref{protonismsr}), while that of $N(1535)$, (\ref{1535ismsr}), 
is smaller.

\subsection{Magnetic moment}
\label{I=1mag}
The isovector magnetic moment is defined as
\begin{eqnarray}
\mu^i_{I=1}=\frac{1}{2}\epsilon^{ijk}
\int d^3x\,x^j \tr(J_V^{k}\tau^3) \times 2\ .
\end{eqnarray}
Here the additional factor of $2$
in the integrand is due to our normalization of the
current.
Substituting (\ref{Ji}) into this expression gives
\begin{eqnarray}
\mu^i_{I=1}
=
- 4\pi^2\kappa\rho^2
\tr(\ba\tau^i\ba^{-1}\tau^3)\ .
\end{eqnarray}
{}For  
the baryon states of $I=J=1/2$,
we can use the identity \cite{ANW}
\begin{eqnarray}
 \bra{B',s'}\tr(\ba\tau^i\ba^{-1}\tau^a)\ket{B,s}
=-\frac{2}{3}(\sigma^i)_{s's}(\tau^a)_{I_3'I_3}\ ,
\label{tita}
\end{eqnarray}
with 
$\sigma^i$ and $\tau^a$ being the Pauli matrices
corresponding to spin and isospin, respectively.
Here we have used the notation
\begin{align}
(\sigma^i)_{s's} =
 \chi_{(s')}^{\dagger}\,\sigma^i\,\chi_{(s)}\ ,
~~~(\tau^C)_{I_3'I_3}=
 \psi^\dagger_{I'_3}\,{\tau^C}\,\psi_{I_3}\ ,
\end{align}
where $\chi_{(s)}$ and $\psi_{I_3}$ are defined as
\begin{align}
\chi_{(1/2)}= \psi_{I_3=1/2}=\left(
\begin{array}{c}
1 \\
0
\end{array}
\right) \ ,~~~
\chi_{(-1/2)}= \psi_{I_3=-1/2}=\left(
\begin{array}{c}
0 \\
1
\end{array}
\right) \ .
\end{align}
Then, we obtain
\begin{eqnarray}
\bra{\,p\!\uparrow\!}\,\mu^i_{I=1}\,\ket{\,p\!\uparrow}
=-\bra{\,n\!\uparrow\!}\,\mu^i_{I=1}\,\ket{\,n\!\uparrow}
=\frac{8\pi^2\kappa}{3}\vev{\rho^2}\,
\delta^{i3}\ ,
\label{pnmu}
\end{eqnarray}
where $\vev{\rho^2}$
is the expectation value of $\rho^2$
for the nucleon states.
Using the wavefunction (\ref{wf}), $\vev{\rho^2}$ is 
obtained as
\begin{align}
 \vev{\rho^2}=
\frac{\int d\rho\,\rho^5 R(\rho)^2}
{\int d\rho\,\rho^3 R(\rho)^2}
=
\frac{\sqrt{5}+2\sqrt{5+N_c^2}}{2N_c}\,\rho^2_{\rm cl} \ .
\label{vevrho}
\end{align}
In the large $N_c$ limit, it can be approximated
by its classical value $\rho_{\rm cl}^2$ in (\ref{rcl1}).
However, for $N_c=3$, (\ref{vevrho}) implies
$\vev{\rho^2}_{n_\rho=0}\simeq 1.62\, \rho_{\rm cl}^2$,
suggesting that the $1/N_c$ corrections are relatively
large for this quantity.

Consequently, we obtain
\begin{eqnarray}
\bra{\,p\!\uparrow\!}\,\mu^i_{I=1}\,\ket{\,p\!\uparrow}
=-\bra{\,n\!\uparrow\!}\,\mu^i_{I=1}\,\ket{\,n\!\uparrow}
=\frac{1+2\sqrt{1+N_c^2/5}}{\sqrt{6}}
\frac{\delta^{i3}}{\Mkk}\ .
\end{eqnarray}
Here we have recovered the $\Mkk$ dependence by dimensional
analysis.
The isovector $g$ factor, defined as in (\ref{gfactor}),
is then
\begin{eqnarray}
g_{I=1}=g_p-g_n=
\frac{4M_N}{\Mkk}\cdot
\frac{1+2\sqrt{1+N_c^2/5}}{\sqrt{6}}
\simeq 7.03
\label{gI1}
\end{eqnarray}
for $N_c=3$, $M_N\simeq 940~{\rm MeV}$,
and (\ref{Mkkkappa}).
The experimental value is
$g_{I=1}|_{\rm exp}\simeq 9.41$ \cite{PDG}
 and the prediction obtained from the Skyrme model
in Ref.~\citen{ANW} is $g_{I=1}|_{\rm ANW}\simeq 6.38$.
If we approximate $\vev{\rho^2}$
by its classical value $\rho_{\rm cl}^2$, we have
$g_{I=1}
\simeq 4.34$, 
which is considerably smaller than the
experimental value.
It is interesting to note that going to
 $\vev{\rho^2}$ from $\rho^2_{\rm cl}$ by 
multiplying by the ratio $(N_c + \sqrt{5}/2)/N_c$
for large $N_c$, as can be seen from (\ref{vevrho}), has
an effect similar to the ``$N_c\rightarrow N_c+2$''
rule discussed in Ref.~\citen{Hong-Rho-Yee-Yi}.
The value of the anomalous magnetic moment obtained using this rule
from the five-dimensional effective spinor 
field theory approach in 
Ref.~\citen{Hong-Rho-Yee-Yi} is close to ours.\footnote{
Again, the same expressions (\ref{pnmu}) and (\ref{gI1}) can be found in 
Ref.~\citen{Hata-Murata-Yamato} but with a different current.
Eq.~(\ref{pnmu}) also agrees with $\Delta\mu^{an}$ in Ref.~\citen{Hong-Rho-Yee-Yi},
if we use the classical value (\ref{rcl1}) for $\vev{\rho^2}$.
}

The magnetic moments for the proton and neutron
(measured in units of the Bohr magneton $\mu_N=1/(2M_N)$)
are given as
\begin{eqnarray}
\mu_p=\frac{g_p}{2}=\frac{1}{4}(g_{I=0}+g_{I=1})\ ,~~
\mu_n=\frac{g_n}{2}=\frac{1}{4}(g_{I=0}-g_{I=1})\ .
\label{mupn}
\end{eqnarray}
If we insert the values (\ref{gI0}) and
(\ref{gI1}), we get
\begin{eqnarray}
\mu_p\simeq 2.18\ ,~~~
\mu_n\simeq -1.34\ ,
\label{mmpn}
\end{eqnarray}
while the experimental values are
$
\mu_p|_{\rm exp}\simeq 2.79$ and
$\mu_n|_{\rm exp}\simeq -1.91$.
Note, however, that since
the $N_c$ dependences of the $g$ factors and the magnetic
moments are
\begin{eqnarray}
 g_{I=0}\sim\cO(1)\ ,~~ g_{I=1}\sim\cO(N_c^2)\ ,
~~ \mu_{I=0}\sim\cO(1/N_c)\ ,~~ \mu_{I=1}\sim\cO(N_c)\ ,
\end{eqnarray}
the contribution of the isoscalar component will
be buried in the $1/N_c$ corrections of the
isovector component in linear combinations such as
(\ref{mupn}). Therefore, it is more meaningful to
consider $g_{I=0}$ and $g_{I=1}$
rather than $g_{p}$ and $g_{n}$ in our analysis.

Let us consider excited baryons.
For spin 1/2 excitations of the baryons, the magnetic moment remains
the same as that of the proton/neutron if $n_\rho=0$ (for example,
$N(1535)$), because it gives the same $\vev{\rho^2}$. On the other hand, 
the Roper excitation $N(1440)$ has $n_\rho=1$ for which, 
with $N_c=3$,
\begin{eqnarray}
\vev{\rho^2}_{n_\rho=1}\simeq 2.37\rho_{\rm cl}^2 \ .
\end{eqnarray}
See Appendix \ref{xx} for the details. 
Substituting it into (\ref{pnmu}) and combining the result with
the value of $\mu_{I=0}$ (which is the same value as that of the
proton/neutron), we obtain, for the Roper excitation,
\begin{eqnarray}
 \mu_{p^*}\simeq 2.99\ ,~~~
 \mu_{n^*}\simeq -2.15\ ,
\end{eqnarray}
measured in units of the Bohr magneton of the nucleon
$1/(2M_N)$.
Our model predicts that the magnetic moment of the Roper is
larger than that of the proton/neutron.

For spin 3/2 baryons such as $\Delta$, we need to reevaluate
the matrix elements of the spin operator $J^i$, (\ref{tita})
and (\ref{vevrho}).
We find (details are described in Appendix \ref{xx2}) 
\begin{align}
&\mu^i_{I=0}(B,J_3=3/2) = 
3 \mu^i_{I=0}(p\uparrow)\ , ~~~
(B=\Delta^{++},\Delta^+,\Delta^0,\Delta^-) 
\label{md++}\\
& 
\mu^i_{I=1}(\Delta^{++},J_3=3/2) 
= \frac{9 c}{5} \mu^i_{I=1}(p\uparrow)\ , 
~~ \mu^i_{I=1}(\Delta^{+},J_3=3/2) = 
\displaystyle\frac{3 c}{5}\mu^i_{I=1}(p\uparrow)\ ,
\nonumber \\
&\mu^i_{I=1}(\Delta^{0},J_3=3/2) 
= -\frac{3 c}{5} \mu^i_{I=1}(p\uparrow)\ , 
~~ \mu^i_{I=1}(\Delta^{-},J_3=3/2) = 
-\displaystyle\frac{9 c}{5} \mu^i_{I=1}(p\uparrow)\ ,
\label{md+}
\end{align}
where $c\equiv \vev{\rho^2}_{l=3}/\vev{\rho^2}_{l=1}\simeq
1.34$.\footnote{This factor was missing in the earlier versions
of the present paper. We thank T.~Ishii for pointing out this error.
}
Therefore we obtain
\begin{eqnarray}
\mu_{\Delta^{++}}\simeq 5.50\ ,~~ 
\mu_{\Delta^{+}}\simeq 2.67\ ,~~
\mu_{\Delta^{0}}\simeq -0.15\ ,~~
\mu_{\Delta^{-}}\simeq -2.97\ ,~~
\label{deltam}
\end{eqnarray}
using (\ref{gI0}) and (\ref{gI1}).
The experimental values for $\Delta^{++}$ and $\Delta^+$ are
\cite{PDG}
\begin{eqnarray}
\mu_{\Delta^{++}}|_{\rm exp}\simeq 3.7 - 7.5\ ,~~~ 
\mu_{\Delta^{+}}|_{\rm exp}\simeq 2.7^{+1.0}_{-1.3}\pm 1.5 \pm 3\ ,
\end{eqnarray}
with which our result (\ref{deltam}) is found to be consistent.
A recent lattice result gives \cite{Cloet:2003jm}
\begin{eqnarray}
\mu_{\Delta^{++}}|_{\rm lattice}\simeq 4.99\ ,~~ 
\mu_{\Delta^{+}}|_{\rm lattice}\simeq 2.49\ ,~~
\mu_{\Delta^{0}}|_{\rm lattice}\simeq 0.06\ ,~~
\mu_{\Delta^{-}}|_{\rm lattice}\simeq -2.45\ ,
~~
\label{Deltalat}
\end{eqnarray}
from which we see that
the agreement with our result (\ref{deltam})
is not very good, particularly for $\mu_{\Delta^0}$.
This is because the contributions of
$\mu_{I=1}\sim\cO(N_c)$ and $\mu_{I=0}\sim\cO(1/N_c)$
roughly cancel each other out for $\Delta^0$
and hence the $1/N_c$ corrections cannot be neglected.
To make a more reasonable comparison, we should
compare $\mu_{I=0}$ and $\mu_{I=1}$ with experimental
or lattice results as explained above. Then,
the average of the magnetic moments in (\ref{Deltalat}),
\begin{eqnarray}
\frac{1}{4}(\mu_{\Delta^{++}}+\mu_{\Delta^{+}}
+\mu_{\Delta^{0}}+\mu_{\Delta^{-}})|_{\rm lattice}\simeq 1.27\ ,
\end{eqnarray}
should be compared with our result for the isoscalar component
$\frac{1}{2}\mu_{I=0}(\Delta)=\frac{3}{2}\mu_{I=0}(p)\linebreak \simeq
1.26 $.
The isovector components are extracted by considering
the differences:
\begin{eqnarray}
&&\frac{1}{3}(\mu_{\Delta^{++}}-\mu_{\Delta^-})|_{\rm lattice}
\simeq 2.48\, ,~
\frac{1}{2}(\mu_{\Delta^{+}}-\mu_{\Delta^-})|_{\rm lattice}
\simeq 2.47\, ,\nn\\
&&(\mu_{\Delta^{0}}-\mu_{\Delta^-})|_{\rm lattice}
\simeq 2.51\, .
\end{eqnarray}
These values are compared with our result
$\frac{3 c}{5}\mu_{I=1}(p)\simeq 2.82$.

\subsection{Axial coupling}
\label{axialcoupling}

As explained in Ref.~\citen{ANW}, the axial coupling $g_A$
of the baryon states of $I=J=1/2$ is given by
\begin{eqnarray}
\int d^3x\,
\langle {B',s'}|J_{A}^{a\,i}|{B,s}\rangle\times 2
=\frac{2}{3}\,g_A\,
(\sigma^i)_{s's}(\tau^a)_{I_3'I_3}\ .
\label{gAdef}
\end{eqnarray}
{}From (\ref{Ji}), the integral of the
axial-vector current becomes
\begin{eqnarray}
\int d^3x\,J_{A}^{a\,i}
=\frac{4}{3}\pi^2\kappa\rho^2
\tr(\ba\tau^i\ba^{-1}\tau^a)
 \int d^3 x\,
\del_j^2 H^A\ .
\label{ddH}
\end{eqnarray}
Although this is an integral of a total derivative,
it does not vanish because of the
term proportional to $1/r$ in (\ref{HA}).
In fact, the integral can be performed
using the Gauss' divergence theorem as
\begin{eqnarray}
 \int d^3x\, \del_j^2 Y_{2n}
= \int_S\, d\vec{S}\cdot\vec{\nabla} Y_{2n}
= 4\pi \lim_{r\ra\infty}
 r^2 \del_r Y_{2n} =\delta^{0n}\ ,
\end{eqnarray}
where $Y_n(r)$ is defined in (\ref{Yukawa})
with $\lambda_0=0$ for $n=0$.
This implies that only
the $n=0$ component of the mesons, that is the pion,
contributes to the integral.
Again using (\ref{tita}), we obtain
\begin{eqnarray}
\int d^3x\,\vev{B',s'|J_{A}^{a\,i}|B,s}
=
-\frac{16\pi\kappa}{9}
\vev{\frac{\rho^2}{k(Z)}}
(\sigma^i)_{s's}(\tau^a)_{I_3'I_3}\ ,
\label{intJA}
\end{eqnarray}
where $\vev{\rho^2/k(Z)}$ is the expectation
value with respect to the spin 1/2 baryon states.
Comparing this with (\ref{gAdef}),
we obtain\footnote{
The sign of this equation is taken to be positive such
that the axial coupling is defined to be positive.
This sign can be flipped if one exchanges the definitions
of ``left'' and ``right'' chiral sectors in 
the two asymptotes $z\rightarrow \pm \infty$,
i.e. the positive sign is a convention.}$^,$\footnote{
This expression, once the classical value $\vev{k(Z)}=1$ is
imposed, is equal to that obtained in 
Ref.~\citen{Hata-Murata-Yamato}. Furthermore, if we use
 $\vev{\rho^2}=\rho^2_{\rm cl}$, it agrees with $g_{A,mag}$ in
Ref.~\citen{Hong-Rho-Yee-Yi}.}
\begin{eqnarray}
 g_A=
\frac{16\pi\kappa}{3}
\vev{\frac{\rho^2}{k(Z)}}\ .
\label{gA2}
\end{eqnarray}

If we approximate $\rho$ and $Z$ by their classical values,
we obtain
\begin{eqnarray}
 g_A\simeq \frac{2N_c}{3\pi}\sqrt{\frac{6}{5}}
\simeq 0.697\ .
\end{eqnarray}
If we use the wavefunction (\ref{wf})
and (\ref{Mkkkappa}) to numerically evaluate the expectation
value for nucleons, we obtain
\begin{eqnarray}
\vev{\frac{\rho^2}{k(Z)}}\simeq 1.05
\, \rho_{\rm cl}^2\ ,
\end{eqnarray}
and 
\begin{eqnarray}
 g_A\simeq 0.734\ .
\end{eqnarray}
The experimental value \cite{PDG} and the prediction obtained from
the Skyrme model in Ref.~\citen{ANW} are
\begin{eqnarray}
 g_A|_{\rm exp}\simeq 1.27\ , 
 ~~~g_A|_{\rm ANW}\simeq 0.61\ .
\end{eqnarray}

For the excited baryons $N(1440)$ (Roper)
with $(n_\rho,n_z)=(1,0)$ and $N(1535)$
with $(n_\rho,n_z)=(0,1)$, by evaluating
 $\vev{\rho^2/k(Z)}$ using their wavefunctions,
we obtain (see Appendix \ref{xx} for details)
\begin{eqnarray}
 g_A^{(N(1440))}\simeq 1.07\ ,~~~
 g_A^{(N(1535))}\simeq 0.380\ .
\end{eqnarray}
We can see that the axial coupling for $N(1440)$ is large
while that for the negative parity baryon
$N(1535)$ is small compared with the proton/neutron.

There is another more direct way to
perform the integral in (\ref{ddH}),
which will be useful in \S \ref{axialradius}.
Using (\ref{HA}) and (\ref{Yn}),
the integrand of (\ref{ddH}) is
\begin{eqnarray}
\del_j^2 H^A
=\left(
\frac{2}{\pi}\frac{1}{k(Z)}-
\sum_{n=1}^\infty\frac{g_{a^n}}{\lambda_{2n}}
\del_Z\psi_{2n}(Z)
\right)\delta^3(\bx-\bX)
-\sum_{n=1}^\infty
 g_{a^n}
\del_Z\psi_{2n}(Z)Y_{2n}(r)\ .
\end{eqnarray}
Following the same logic as in (\ref{NBsum}),
we can show that
\begin{eqnarray}
\sum_{n=1}^\infty \frac{g_{a^n}}{\lambda_{2n}}
\psi_{2n}(Z)
=\psi_0(Z)\ ,
\label{sumga1}
\end{eqnarray}
where $\psi_0(z)=\frac{2}{\pi}\arctan z$,
using (\ref{gvga2}) and (\ref{complete}).
This relation implies
\begin{eqnarray}
\sum_{n=1}^\infty \frac{g_{a^n}}{\lambda_{2n}}
\del_Z\psi_{2n}(Z)
=\frac{2}{\pi}\frac{1}{k(Z)}\ ,
\label{sumga2}
\end{eqnarray}
and hence
\begin{eqnarray}
\del_j^2 H^A
=-\sum_{n=1}^\infty
 g_{a^n}
\del_Z\psi_{2n}(Z)Y_{2n}(r)\ .
\end{eqnarray}
Then, we can perform the integral as
\begin{eqnarray}
\int d^3 x\,
\del_j^2 H^A
=\int_0^\infty dr \,r
\sum_{n=1}^\infty g_{a^n}
\del_Z\psi_{2n}(Z) e^{-\sqrt{\lambda_{2n}}r}
=\frac{2}{\pi}\frac{1}{k(Z)}\ ,
\label{intddH}
\end{eqnarray}
using (\ref{sumga2}). From this,
it is easy to rederive (\ref{intJA}).

\subsection{Goldberger-Treiman relation}

The axial coupling obtained in the previous
subsection is related to the $\pi NN$ coupling
$g_{\pi NN}$ by the Goldberger-Treiman relation
\begin{eqnarray}
 g_A=\frac{f_\pi g_{\pi NN}}{M_N}\ .
\label{GT}
\end{eqnarray}
We can derive this relation in our context
following the argument given in Ref.~\citen{ANW} for
the Skyrme model.

It is argued in Ref.~\citen{ANW} that the pion field behaves
asymptotically as
\begin{eqnarray}
 \vev{\Pi^a(x)}\simeq
-\frac{g_{\pi NN}}{8\pi M_N}\,\frac{x^i}{r^3}\,
\vev{\sigma^i\tau^a}
\label{pi1}
\end{eqnarray}
in the presence of a nucleon.
Here, $g_{\pi NN}$ is the $\pi NN$ coupling
and the expectation value is taken for the nucleon.
In the present case, (\ref{AzGc}) provides us with
information on the asymptotic pion field.
The pion field can be read from the
$n=0$ component of (\ref{H1}) substituted
in (\ref{AzGc}). We find
\begin{eqnarray}
A_z^G\simeq
\Pi(x)\phi_0(z)+\cdots
\end{eqnarray}
with
\begin{eqnarray}
 \Pi(x)=\frac{\sqrt{\kappa\pi}}{2}
\ba\tau^i\ba^{-1}
\frac{\rho^2}{k(Z)}\frac{x^i}{r^3}\ ,
\end{eqnarray}
which implies
\begin{eqnarray}
\vev{\Pi^a(x)}\simeq
-\frac{\sqrt{\kappa\pi}}{3}
\vev{\frac{\rho^2}{k(Z)}}\frac{x^i}{r^3}
\vev{\sigma^i\tau^a}\ .
\label{pi2}
\end{eqnarray}
Comparing (\ref{pi1}) and (\ref{pi2}),
we obtain
\begin{eqnarray}
 \frac{g_{\pi NN}}{M_N}=
\frac{8\pi\sqrt{\kappa\pi}}{3}\vev{\frac{\rho^2}{k(Z)}}
=\frac{1}{2}\sqrt{\frac{\pi}{\kappa}}g_A
=\frac{g_A}{f_\pi}\ ,
\end{eqnarray}
where we have used (\ref{const}) and (\ref{gA2}).
This is nothing but the Goldberger-Treiman relation
(\ref{GT}).

\subsection{Axial radius}
\label{axialradius}

We define $\rho_A(r)$ as a function
proportional to the expectation value of the integrand of
(\ref{intddH}) with the normalization
$\int_0^\infty dr\,\rho_A(r)=1$,
\begin{eqnarray}
\rho_A(r)\equiv
\vev{
r\sum_{n=1}^\infty g_{a^n}
\del_Z\psi_{2n}(Z)
e^{-\sqrt{\lambda_{2n}}r}}\Big/
\vev{
\frac{2}{\pi}\frac{1}{k(Z)}}\ .
\end{eqnarray}
The axial radius is obtained as
\begin{align}
\vev{r^2}_{A}\equiv
 \int_0^\infty dr \,r^2\rho_A(r)
=\vev{F_A(Z)}/\vev{k(Z)^{-1}}\ ,
\end{align}
where
\begin{eqnarray}
F_A(z)\equiv
3\pi \sum_{n=1}^\infty
\frac{ g_{a^n}}{\lambda_{2n}^2}
\del_z\psi_{2n}(z)\ .
\label{fA}
\end{eqnarray}
{}From (\ref{psi1}) and (\ref{sumga1}),
this function satisfies the differential equation
\begin{eqnarray}
 \del_z(k(z)F_A(z))
=-3\pi h(z)\psi_0(z)\ .
\label{fAd}
\end{eqnarray}
Integrating this equation, we obtain
\begin{eqnarray}
k(z)F_A(z)=f_0-3\pi\int_0^z dz'h(z')\psi_0(z')\ ,
\label{fA2}
\end{eqnarray}
where $f_0$ is a constant.
In order to fix $f_0$, we use the identity
\begin{eqnarray}
 \int dz\,F_A(z)=0\ ,
\label{intfAk}
\end{eqnarray}
which follows from (\ref{fA}).
Substituting (\ref{fA2}) into (\ref{intfAk}), we obtain
\begin{eqnarray}
f_0=3\int_{-\infty}^\infty dz\,
\frac{1}{k(z)}\int_0^zdz'h(z')\psi_0(z')\simeq 7.82\ .
\end{eqnarray}
Therefore, if we approximate the expectation
value by its classical value, we obtain
\begin{eqnarray}
 \vev{r^2}_{A}=\vev{F_A(Z)}/\vev{k(Z)^{-1}}\simeq F_A(0)=f_0
\simeq 7.82/\Mkk^2\simeq (0.582~{\rm fm})^2\ .
\end{eqnarray}
If we numerically evaluate the expectation value
of $F_A(Z)$ using the wavefunction (\ref{wf}),
we obtain
\begin{eqnarray}
 \vev{r^2}_{A}^{1/2}\simeq 0.537~{\rm fm}\ .
\label{pnar}
\end{eqnarray}
The experimental value is 
$\vev{r^2}_{A}^{1/2}|_{\rm exp}\simeq 0.674~{\rm fm}$
\cite{BoAvBrBu}.$^,$\footnote{This value is obtained by
applying the formula 
$\langle r^2 \rangle_A=-6d/(dk^2)\log g_A(k^2)|_{k^2=0}$
to the axial form factor $g_A(k^2)$ in Ref.~\citen{BoAvBrBu},
which is fitted by a dipole.
}

We can compute the axial radius for excited baryons
in the same manner as for the other quantities evaluated before.
The computation of the axial radius is independent of
the profile of $R(\rho)$ in the wavefunction;
thus, for the $N(1440)$ (Roper) with
$(n_\rho,n_z)=(1,0)$, it gives the same result as that for the
proton/neutron (\ref{pnar}). 
For the $N(1535)$ with $(n_\rho,n_z)=(0,1)$, 
we obtain (see Appendix \ref{xx} for details)
\begin{eqnarray}
 \vev{r^2}_A^{1/2}\simeq 0.435~{\rm fm}\ .
\end{eqnarray}
This is smaller than the axial radius of the
proton/neutron. 


\section{Form factors}

In this section, we compute the form factors of
spin 1/2 baryons associated with the currents $\cJ_{V,A}^\mu$
obtained in the previous section.
To this end, we first give a brief review of how to compute
the form factors from the matrix elements of the currents.
The present model enables us to compute the matrix elements 
easily with the tools formulated in \S \ref{seccurrents}.

\subsection{Formalism}

Let us first
consider the matrix elements of a vector current
for a baryon of spin 1/2,
\begin{align}
\langle \bp',B',s'|\cJ_V^{C\mu}(0)
|\bp,B,s\rangle \,.
\end{align}
Here $|\bp,B,s\rangle$ and $|\bp',B',s'\rangle$ 
denote the initial
and final states of the baryon under consideration.
In the present section, we focus on the case where $B$ and $B'$
have the same $n_\rho$ and $n_z$ while the isospins $I_3$ and
$I'_3$ may be different.
The states are normalized as in (\ref{normalize}).
The most general form of the matrix elements
consistent with the symmetries 
and conservation of the current is
\begin{align}
 \langle \bp',B',s'|\cJ_V^{C\mu}(0)|\bp,B,s\rangle=i(2\pi)^{-3}
\frac{(\tau^C)_{I_3'I_3}}{2}
\,\ol u(\bp',s')\Gamma^\mu_{(C)}(p',p)\, u(\bp,s)
\label{JGamma}
\end{align}
with
\begin{align}
& \ol u(\bp',s')\Gamma^\mu_{(0)}(p',p)\, u(\bp,s) 
= \,\ol u(\bp',s')\left[
\gamma^\mu \wh F_1(k^2)-\frac{1}{2m_B}
\sigma^{\mu\nu}k_\nu\,\wh F_2(k^2)
\right] u(\bp,s) 
\ ,
\label{diracpauli1}\\
& \ol u(\bp',s')\Gamma^\mu_{(1,2,3)}(p',p)\, u(\bp,s) 
= \,\ol u(\bp',s')\left[
\gamma^\mu F_1(k^2)-\frac{1}{2m_B}
\sigma^{\mu\nu}k_\nu\,F_2(k^2)
\right] u(\bp,s) 
\ ,
\label{diracpauli2}
\end{align}
where $k=p-p'$ and $m_B$ is the baryon mass.
On the right-hand side of (\ref{JGamma}), no summation
is taken for the index $C$.
$F_1(k^2)$ and $F_2(k^2)$ are the scalar functions of
$k^2$ called the Dirac and Pauli form factors,
respectively, whose
dependence on $n_\rho$ and $n_z$ is indicated implicitly.
They can be computed by evaluating the matrix elements of
the current.
$u(\bp,s)$ and $u(\bp',s')$ are the Dirac spinors associated with
the initial and final states of the baryon of mass $m_B$, respectively.
The normalization condition\footnote{
With this normalization, 
we assign $\sqrt{2p^0}\,u(\bp,s)$ and 
$\sqrt{2p^0}\,\ol u(\bp,s)$
to an incoming and outgoing external line, respectively,
in the computation of the Lorentz invariant matrix elements.}
is given by
\begin{align}
\ol u(\bp,s') u(\bp,s)=\delta_{s's}\,\frac{m_B}{p^0} \ .
\end{align}

It is useful to write the matrix elements in the Breit frame with
$\bp=-\bp'=\bk/2$ and $E=E'=\sqrt{m_B^2+\vec k^2/4}$:
\begin{align}
 \langle -\frac{\bk}{2},B',s'
|\wh J_V^{0}(0)
\,|\frac{\bk}{2},B,s\rangle
&=
(2\pi)^{-3}\,\frac{\delta_{I_3'I_3}}{2}
\delta_{s's}\,\frac{m_B}{E}\,
\wh G_E(\bk^2) \ ,
\nn\\
 \langle -\frac{\bk}{2},B',s'
|\wh J_V^{j}(0)
|\,\frac{\bk}{2},B,s\rangle
&=
(2\pi)^{-3}\,\frac{\delta_{I_3'I_3}}{2}
\,\frac{i}{2E}\epsilon_{jla}\,k_l\left(\sigma^a\right)_{s's}
\wh G_M(\bk^2) \ ,
\nn\\
\langle -\frac{\bk}{2},B',s'
|J_V^{c0}(0)
\,|\frac{\bk}{2},B,s\rangle
&=
(2\pi)^{-3}\,\frac{\left(\tau^c\right)_{I_3'I_3}}{2}
\delta_{s's}\,\frac{m_B}{E}\,
G_E(\bk^2) \ ,
\nn\\
 \langle -\frac{\bk}{2},B',s'
|J_V^{cj}(0)
|\,\frac{\bk}{2},B,s\rangle
&=
(2\pi)^{-3}\,\frac{\left(\tau^c\right)_{I_3'I_3}}{2}
\,\frac{i}{2E}\epsilon_{jla}\,k_l\left(\sigma^a\right)_{s's}
G_M(\bk^2) \ .
\label{breit}
\end{align}
Here $G_{E,M}(\bk^2)$ are 
the Sachs form factors, related to the Dirac and
Pauli form factors by
\begin{align}
&\wh G_E(\bk^2)
=\wh F_1(\bk^2)-\frac{\bk^2}{4m_B^2}\,\wh F_2(\bk^2)
\ ,~~
\wh G_M(\bk^2)
=\wh F_1(\bk^2)+\wh F_2(\bk^2)
\ ,
\nn\\
&G_E(\bk^2)
=F_1(\bk^2)-\frac{\bk^2}{4m_B^2}\,F_2(\bk^2)
\ ,~~
G_M(\bk^2)
=F_1(\bk^2)+F_2(\bk^2)
\ .
\label{sachs:dp}
\end{align}
The formulae of the Dirac spinor needed for this manipulation
are summarized in Appendix \ref{dirac}.
The Sachs form factors can be obtained by evaluating the left-hand side
of (\ref{breit}) using the baryon
wavefunctions given in
 \S \ref{rev:model}
and Appendix \ref{ap:wv}.

The electromagnetic form factors
are defined by considering the matrix elements of the
electromagnetic current,
\begin{eqnarray}
 J_{\rm em}^\mu=J_V^{a=3,\mu}+\frac{1}{N_c}\wh J^\mu
\ .
\end{eqnarray}
Then, for the states with $I_3=+1/2$ (p) and
 $I_3=-1/2$ (n), the Sachs form factors
 associated with the electromagnetic current
are given by
\begin{align}
&G_{E,M}^{\rm p}(\bk^2)=\half
\left(+G_{E,M}(\bk^2)+\frac{1}{N_c}\wh G_{E,M}(\bk^2)
\right)\ ,~~~(\mbox{for $I_3=+1/2$})\nn\\
&G_{E,M}^{\rm n}(\bk^2)=\half
\left(-G_{E,M}(\bk^2)+\frac{1}{N_c}\wh G_{E,M}(\bk^2)
\right)\ ,~~~(\mbox{for $I_3=-1/2$})
\label{GEM}
\end{align}
respectively.
 $G_E^{\rm p,n}$ and $G_M^{\rm p,n}$
are called the electric and magnetic Sachs form factors, and
their Fourier transformation
provides the distribution of the
electric charge density and the magnetic current
density, respectively.
(See Refs.~\citen{Me,ArRoZa,PePuVa} for reviews.)

Next we study
the axial form factor associated with the axial 
current $\cJ_A^{\mu}(x)$.
As in the vector current case, we consider 
the matrix elements of the axial current 
for a spin 1/2 baryon,
\begin{align}
\langle \bp',B',s'|\cJ_A^{C\mu}(0)
|\bp,B,s \rangle \ .
\end{align}
The matrix elements consistent with the symmetries
can be written in terms of
the axial form factor $g_A(k^2)$ and
the induced pseudoscalar form factor $g_P(k^2)$ as
\begin{align}
\langle \bp',B',s'|\cJ_A^{C\mu}(0)|\bp,B,s \rangle
&=(2\pi)^{-3}
\frac{(\tau^C)_{I_3'I_3}}{2}\,
\ol u(\bp',s')\Gamma^\mu_{A(C)}(p',p)\,
 u(\bp,s)
\label{AA0}
\end{align}
with
\begin{align}
&\ol u(\bp',s') \Gamma^\mu_{A(0)}(p',p)
u(\bp,s)
=
\ol u(\bp',s')\Big[
i\gamma_5\gamma^\mu\, \wh g_A(k^2)
+\frac{1}{2m_B}k^\mu\gamma_5\,\wh g_P(k^2)\Big] u(\bp,s)\ ,
\label{whgAgP}\\
&\ol u(\bp',s') \Gamma^\mu_{A(1,2,3)}(p',p)
u(\bp,s)
=
\ol u(\bp',s')\Big[
i\gamma_5\gamma^\mu\, g_A(k^2)
+\frac{1}{2m_B}k^\mu\gamma_5\,g_P(k^2)\Big] u(\bp,s)\ .
\label{gAgP}
\end{align}
On the right-hand side of (\ref{AA0}), no summation is taken for
the index $C$.
The current conservation law yields
\begin{align}
 \wh g_P(k^2)=\frac{4m_B^2}{k^2}\,\wh g_A(k^2) \ ,~~~
 g_P(k^2)=\frac{4m_B^2}{k^2}\,g_A(k^2) \ .
\label{h-g}
\end{align}
In the nonrelativistic limit, the spatial component becomes
\begin{align}
 \langle \bp',B',s'
|\wh J_A^{j}(0)|\bp,B,s \rangle\simeq&
-(2\pi)^{-3}
(\sigma^a)_{s's}\,
\frac{\delta_{I'_3I_3}}{2}\left(\delta_{ja}-\frac{k_jk_a}{k^2}\right)
\wh g_A(k^2) \ ,
\nn
\\
 \langle \bp',B',s'
|J_A^{cj}(0)|\bp,B,s \rangle\simeq&
-(2\pi)^{-3}
(\sigma^a)_{s's}\,
\frac{(\tau^c)_{I'_3I_3}}{2}\left(\delta_{ja}-\frac{k_jk_a}{k^2}\right)
 g_A(k^2) \ .
\label{m:axial}
\end{align}
Again, the axial form factor $g_A(k^2)$ can be computed by evaluating the
matrix elements in  (\ref{m:axial})
 from the baryon wavefunctions 
given in \S \ref{rev:model}
and Appendix \ref{ap:wv}.

In order to derive the form factors from the above formalism,
it is useful to perform the Fourier transformation of the currents
defined by
\begin{eqnarray}
\wt \cJ^\mu(\vec k)=\int d^3 x \,e^{-i\vec k\cdot \vec x}\cJ^\mu(x) \ .
\end{eqnarray}
Using the explicit form of the current (\ref{J0U1}), (\ref{JiU1}),
(\ref{J0}), and (\ref{Ji}), 
it is not difficult to show that
\begin{align}
\wt{\wh J_V}^0(\bk)&=e^{-i\bk\cdot\bX}\,\frac{N_c}{2}
\sum_{n\ge 1}\frac{g_{v^n}\psi_{2n-1}(Z)}{\bk^2+\lambda_{2n-1}} \ ,
\label{wtwhJ0}
\\
\wt{\wh J_V}^j(\bk)&=e^{-i\bk\cdot\bX}\,\frac{N_c}{2}
\left(\frac{P_X^j-k_j/2}{M_0}+\frac{i}{16\pi^2\kappa}
\,\epsilon_{jla}\,k_l\,J_a\right)
\sum_{n\ge 1}\frac{g_{v^n}\psi_{2n-1}(Z)}{\bk^2+\lambda_{2n-1}}+\cdots \ ,
\label{wtwhJj}
\end{align}
\begin{align}
 \wt J_V^{c0}(\bk)&=e^{-i\bk\cdot\bX}\left[
I_c-i\,2\pi^2\kappa\rho^2\tr\left(\tau^c\ba\tau^a\ba^{-1}\right)
\epsilon_{jla}\,k_l\,\frac{P_X^j}{M_0}\right]
\sum_{n\ge 1}\frac{g_{v^n}\psi_{2n-1}(Z)}{\bk^2+\lambda_{2n-1}}+\cdots \ ,
\label{wtJ0}\\
 \wt J_V^{cj}(\bk)&=e^{-i\bk\cdot\bX}(-i)\,2\pi^2\kappa\rho^2\,
\epsilon_{jla}k_l\tr\left(\tau^c\ba\tau^a\ba^{-1}\right)
\sum_{n\ge 1}\frac{g_{v^n}\psi_{2n-1}(Z)}{\bk^2+\lambda_{2n-1}}+\cdots \ ,
\label{wtJj}
\end{align}
\begin{align}
 \wt{\wh J_A}^j(\bk)&=e^{-i\bk\cdot\bX}\,
(-1)\frac{N_c}{32\pi^2\kappa}
\,J_a\left(\delta_{aj}-\frac{k_ak_j}{\bk^2}\right)
\sum_{n\ge 1}\frac{g_{a^n}\partial_Z\psi_{2n}(Z)}{\bk^2+\lambda_{2n}}+\cdots \ ,
\\
 \wt J_A^{cj}(\bk)&=e^{-i\bk\cdot\bX}\,
2\pi^2\kappa\rho^2\tr\left(\tau^c\ba\tau^a\ba^{-1}\right)
\left(\delta_{aj}-\frac{k_ak_j}{\bk^2}\right)
\sum_{n\ge 1}\frac{g_{a^n}\partial_Z\psi_{2n}(Z)}{\bk^2+\lambda_{2n}}
+\cdots \ ,
\end{align}
where we have used (\ref{mom}), (\ref{chi}), and (\ref{iso}).
Here `$\cdots$' denotes the terms that are odd with
respect to $Z$;\footnote{More precisely,
$\wt{\wh J_V}^j$ includes a term that is even in $Z$
and proportional to $\dot{Z}$. We discard this term, since
it is negligible for large $N_c$ and $\lambda$.
}
they do not contribute to the result for the matrix elements.
Useful formulas here are
\begin{align}
& \int d^3x\,e^{-i\bk\cdot\bx}\,Y_n(|\bx-\bX|)
=-e^{-i\bk\cdot\bX}\,\frac{1}{\bk^2+\lambda_n} \ ,\\
& \int d^3x\,e^{-i\bk\cdot\bx}\, H^A(Z,|\bx-\bX|)
=-e^{-i\bk\cdot\bX}\,
\frac{1}{\bk^2}\sum_{n=1}^\infty
\frac{g_{a^n}\del_Z\psi_{2n}(Z)}{\bk^2+\lambda_{2n}}\ .
\end{align}
The latter can be shown by using (\ref{sumga2}).
In addition,
it is important to note that the operator ordering in the
Fourier transformation of the currents is fixed uniquely by 
the requirement
\begin{align}
 \left(\wt\cJ_{V,A}^\mu(\bk)\right)^{\dagger}=
\wt\cJ_{V,A}^\mu(-\bk) \ ,
\end{align}
which is equivalent to the Hermiticity of the current in the 
$x$-representation.

Using these expressions for the Fourier transformed currents
together with the relation
$\langle \vec p\,'|e^{-i\bk\cdot\vec X}|\vec p \rangle =
\delta^3(\bk-\vec p+\vec p\,')$,
the matrix elements can be calculated as
\begin{eqnarray}
\langle \vec p\,',B',s'|\cJ^\mu(0)|\vec p,B,s\rangle
=\int \frac{d^3 k}{(2\pi)^3}
\langle \vec p\,',B',s'|\wt\cJ^\mu(\bk)|\vec p,B,s\rangle\ .
\end{eqnarray}

\subsection{Form factors}

Comparing (\ref{wtwhJ0})--(\ref{wtJj}) with (\ref{breit}),
we find
\begin{align}
&\wh G_E(\bk^2)=N_c
\sum_{n\ge 1}\frac{g_{v^n}\langle\psi_{2n-1}(Z)\rangle} 
{\bk^2+\lambda_{2n-1}} \ ,
\quad
\wh G_M(\bk^2)=N_c\,\frac{g_{I=0}}{2}\,
\sum_{n\ge 1}\frac{g_{v^n}\langle\psi_{2n-1}(Z)\rangle}
{\bk^2+\lambda_{2n-1}} \ ,
\label{GhEM}
\\
\nn\\
&G_E(\bk^2)=
\sum_{n\ge 1}\frac{g_{v^n}\langle\psi_{2n-1}(Z)\rangle} 
{\bk^2+\lambda_{2n-1}} \ ,
\quad
G_M(\bk^2)=
\frac{g_{I=1}}{2}
\sum_{n\ge 1}\frac{g_{v^n}\langle\psi_{2n-1}(Z)\rangle}
{\bk^2+\lambda_{2n-1}} \ ,
\label{GEM-2}
\end{align}
with
\begin{align}
g_{I=0}=\frac{m_B}{M_0} \ ,~~~
g_{I=1}=\frac{32\pi^2\kappa m_B}{3}\langle \rho^2\rangle  
\end{align}
being the isoscalar and isovector $g$ factors
of the baryon $B$, as derived in \S\S \ref{I=0mag} and
\ref{I=1mag}, respectively.
Here we have used the formula (\ref{tita}).
We note also
\begin{align}
 \wh G_E(0)=N_c \ ,~~~
 G_E(0)=1 \ .
\end{align}
Using (\ref{GEM}), the electric and magnetic Sachs form
factors for nucleons and their excited states with $I=J=1/2$
are obtained as
\begin{align}
&G_E^{\rm p}(\bk^2)=
\sum_{n\ge 1}\frac{g_{v^n}\langle\psi_{2n-1}(Z)\rangle}
{\bk^2+\lambda_{2n-1}} \ ,~~G_E^{\rm n}(\bk^2)=0\ ,
\nn\\
&G_M^{\rm p,n}(\bk^2)=\frac{g_{\rm p,n}}{2}\,
\sum_{n\ge 1}
\frac{g_{v^n}\langle\psi_{2n-1}(Z)\rangle}{\bk^2+\lambda_{2n-1}} \ ,
\end{align}
where 
\begin{align}
g_{\rm p}= \frac{1}{2}\left(g_{I=0}+g_{I=1}\right) \ ,
~~
g_{\rm n}= \frac{1}{2}\left(g_{I=0}- g_{I=1}\right) 
\end{align}
are the $g$ factors of the nucleons (and their excitations).
It follows that they satisfy the relation
\begin{align}
 G_E^{\rm p}(\bk^2)=\frac{2}{g_{\rm p}}G_M^{\rm p}(\bk^2)
=\frac{2}{g_{\rm n}}G_M^{\rm n}(\bk^2)
=\sum_{n\ge 1}
\frac{g_{v^n}\langle\psi_{2n-1}(Z)\rangle}{\bk^2+\lambda_{2n-1}} 
~ \ ,~~~
G_E^{\rm n}(\bk^2)=0 \ .
\label{propto}
\end{align}

\begin{figure}
\centerline{\includegraphics{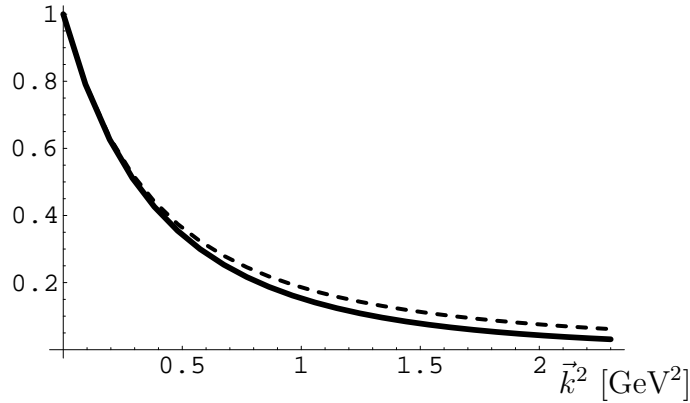}
\put(0,0){\makebox(0,0){$\bk^2$ [$\mbox{GeV}^2$]}}}
\vspace*{3mm}
\caption{\footnotesize{
A plot of $F(\bk^2,0)$, which is equal to $G_E^{\rm p}(\bk)$,
with $\vev{\psi_{2n-1}(Z)}$ approximated by
the classical value $\psi_{2n-1}(0)$.
Our result (solid line) reproduces the dipole behavior (dotted line)
in (\ref{dip})
with $\Lambda^2=0.758~\mbox{GeV}^2$.
}}
\label{plot}
\end{figure}

Experimentally, the Sachs form factors for the proton
and neutron are known to be well described as 
\begin{align}
 G_E^{\rm p}(\bk^2)=\frac{G_M^{\rm p}(\bk^2)}{\mu_{\rm p}}
=\frac{G_M^{\rm n}(\bk^2)}{\mu_{\rm n}}
=\left(1+\frac{\bk^2}{\Lambda^2}\right)^{-2}
 \ ,~~~
G_E^{\rm n}(\bk^2)=0 \ ,
\label{dipole}
\end{align}
with $\Lambda^2=0.71~\mbox{GeV}^2$.
That is, the three form factors $G_E^{\rm p}$, $G_M^{\rm p}$, and
$G_M^{\rm n}$ are proportional to each other and characterized 
by the dipole behavior. 
{}Furthermore, the electric charge density of the 
neutron can be well approximated to be flat.
It turns out that our result (\ref{propto}) is in accord with
these experimental results.
In particular, the infinite sum in (\ref{propto}) 
can be approximated by a single dipole factor,
showing the agreement as a function of $\bk^2$. 
To see this, we expand our result 
(\ref{propto}) as a Taylor series in $\bk^2$, 
\begin{eqnarray}
\sum_{n\ge 1}
\frac{g_{v^n}\langle\psi_{2n-1}(Z)\rangle}{\bk^2+\lambda_{2n-1}} 
= \vev{f_0(Z)} - \vev{f_1(Z)} \bk^2 + \vev{f_2(Z)} (\bk^2)^2
 -\vev{f_3(Z)} (\bk^2)^3 
+\cdots \  ,
~~
\end{eqnarray}
where the coefficients are obtained as the expectation values of
\begin{eqnarray}
f_k(Z)\equiv\sum_{n\ge 1} \frac{g_{v^n} 
{\psi_{2n-1}(Z)}}{(\lambda_{2n-1})^{k+1}} \ .
\end{eqnarray}
As seen in (\ref{NBsum}), we have $\vev{f_0(Z)}=1$.
Let us evaluate these coefficients
using the classical approximation
$\vev{\psi_{2n-1}(Z)}\simeq\psi_{2n-1}(0)$. The first nontrivial coefficient,
$f_1(0)$, is in fact equal to $F_0/6$ given in (\ref{rI0-2}). 
The higher coefficients can be obtained in the same
way as $F_0$, as they satisfy the recursive relation
\begin{eqnarray}
-\del_z(k(z) \del_z f_{k}(z)) = h(z)f_{k-1}(z)\ ,~~
f_{k}(z)=f_{k}(-z)\ ,~~f_k(\pm\infty)=0\ .
\label{recur}
\end{eqnarray}
We thus obtain
\begin{eqnarray}
 f_1(0)=2.38 \; , \; f_2(0)=4.02 \; , \;
 f_3(0)=6.20 \; , \; f_4(0)=9.35 \; , \; 
f_5(0)=14.0 \; , \; \cdots \; .
\label{expcoef}
\end{eqnarray}
On the other hand, the dipole expression (\ref{dipole}) can be expanded 
using $\Lambda^2 = 0.758~{\rm GeV}^2$ as
\begin{eqnarray}
 \left(1+\frac{\bk^2}{\Lambda^2}\right)^{-2}
= 1 -2.38 \bk^2 + 4.24 (\bk^2)^2 - 6.71 (\bk^2)^3
+9.97 (\bk^2)^4 - 14.2 (\bk^2)^5 + \cdots \; ,
~~
\label{dip}
\end{eqnarray}
where we have chosen 
the parameter $\Lambda^2$ in such a way that the first
coefficient reproduces the $f_1(0)$ of (\ref{expcoef})
with the
unit $M_{\rm KK}=949~{\rm MeV}=1$. 
We find a good agreement for the latter coefficients,
suggesting that our form factors (\ref{propto})
exhibit the dipole behavior indicated by experiments.
It is also useful to note that the relation (\ref{recur})
implies that the function
\begin{eqnarray}
F(\bk^2,z)\equiv \sum_{n\ge 1}\frac{g_{v^n}\psi_{2n-1}(z)}
{\bk^2+\lambda_{2n-1}}
\label{FF}
\end{eqnarray}
satisfies
\begin{eqnarray}
\del_z(k(z)\del_z F(\bk^2,z))=\bk^2 h(z) F(\bk^2,z)\ ,~~
F(\bk^2,z)=F(\bk^2,-z)\ ,~~F(\bk^2,\pm\infty)=1\ .
~~
\end{eqnarray}
The solution of this equation evaluated at $z=0$ is
plotted in Fig. \ref{plot}.

The electric and magnetic charge radii can be computed from the first
coefficient of the form factor expanded in powers of $\bk^2$.
{Namely, they are given by
\begin{eqnarray}
\vev{r^2}_{E,M} = -6 \frac{d}{d \bk^2} 
\log G_{E,M}(\bk^2) \biggm|_{\bk^2=0} \ ,
\label{logG}
\end{eqnarray}
except for the neutron charge radius, which is defined by
\begin{eqnarray}
\vev{r^2}_{E,\rm n} = -6 \frac{d}{d \bk^2} 
G_E^{\rm n}(\bk^2) \biggm|_{\bk^2=0} \ .
\end{eqnarray}

Since all the form factors are proportional to each other (except for
 $G_E^{\rm n}$, which vanishes) as in (\ref{propto}), 
we conclude that
\begin{eqnarray}
 \vev{r^2}_{M,\rm p} 
=\vev{r^2}_{M,\rm n} 
=\vev{r^2}_{E,\rm p} \ , \quad 
 \vev{r^2}_{E,\rm n}=0 \ .
\end{eqnarray}
It is easy to check that the definition of the charge radii
in (\ref{logG}) is consistent with our previous calculation
in \S \ref{secstatic},
and using our result for the electric charge radius of the proton
(\ref{protonismsr}), we obtain
\begin{eqnarray}
 \vev{r^2}_{E,\rm p}^{1/2}
= \vev{r^2}_{M,\rm p}^{1/2}
=  \vev{r^2}_{M,\rm n}^{1/2}
 \simeq ~ 0.742 ~ {\rm fm} \ .
\label{radii}
\end{eqnarray}
The values observed in experiments are 
$\vev{r^2}_{E,\rm p}^{1/2}|_{\rm exp}\simeq ~0.875~{\rm fm}$
\cite{PDG},
$\vev{r^2}_{M,\rm p}^{1/2}|_{\rm exp}\simeq \linebreak ~0.855~{\rm fm}$
\cite{HyJa}, and 
$\vev{r^2}_{M,\rm n}^{1/2}|_{\rm exp}\simeq ~ 0.873~{\rm fm}$,
\cite{Ku}
which are reasonably close to our result.

The axial form factors are given by
\begin{align}
&\wh g_A(\bk^2)=
\frac{N_c}{32\pi^2\kappa} 
\sum_{n\ge 1}\frac{g_{a_n}\langle\partial_Z\psi_{2n}(Z)\rangle}
{\bk^2+\lambda_{2n}} \ ,
\quad 
g_A(\bk^2)
=\frac{8\pi^2\kappa}{3}\,\langle\rho^2\rangle
\sum_{n\ge 1}\frac{g_{a_n}\langle\partial_Z\psi_{2n}(Z)\rangle}
{\bk^2+\lambda_{2n}} \ .
\label{gA}
\end{align}
Note that the values at $\bk^2=0$ are obtained by
using (\ref{sumga2}) as
\begin{align}
\wh g_A(0)
=\frac{N_c}{16\pi^3\kappa}\Big\langle \frac{1}{k(Z)}\Big\rangle
\ ,
\quad
g_A(0)
=\frac{16\pi\kappa}{3}
\Big\langle \frac{\rho^2}{k(Z)}\Big\rangle
 \ .
\label{gA0}
\end{align}
The latter reproduces (\ref{gA2}).

It is empirically known that
the axial form factor $g_A(k^2)$ can also
be well fitted by a dipole profile.
Using the same technique as above, 
the Taylor expanded axial form factor with the classical
approximation
$\vev{\del_Z\psi_{2n}(Z)}\simeq \del_z\psi_{2n}(0)$
is found to take the form
\begin{eqnarray}
\frac{g_A(\vec k^2)}{g_A(0)}\simeq
1-1.30 \vec k^2
+1.09 (\vec k^2)^2- 0.770  (\vec k^2)^3
+0.511 (\vec k^2)^4- 0.331  (\vec k^2)^5+\cdots\ .
\end{eqnarray}
These coefficients are close to those obtained from 
the dipole profile
\begin{eqnarray}
&&\left(1+\frac{\vec k^2}{M_A^2}\right)^{-2}\nn\\
&=&1-1.13 \vec k^2
+0.958 (\vec k^2)^2- 0.721  (\vec k^2)^3
+0.510 (\vec k^2)^4- 0.345  (\vec k^2)^5+\cdots\ ,
\end{eqnarray}
with $M_A\simeq 1.26~{\rm GeV}$.

\subsection{Cubic coupling}
The form factors computed above
are composed of an infinite tower of poles that correspond
to the vector and axial-vector meson exchange. From the residues, 
we can extract information on 
cubic couplings among
baryons and (axial-)vector mesons.

We assume that there exist cubic couplings of the form
\begin{align}
 \cL_{\rm int}^v=&\sum_{n\ge 1}\left(
\wh g_{v^nBB}\,
\wh v_\mu^n\,\ol Bi\gamma^\mu\frac{\tau^0}{2}B
+
g_{v^nBB}\,
v_\mu^{na}\,\ol Bi\gamma^\mu\frac{\tau^a}{2}B
\right) \nn\\
+\frac{1}{4m_B}&\sum_{n\ge 1}
\left(
\wh h_{v^nBB}\left(\partial_\mu\wh v_\nu^n
-\partial_\nu\wh v_\mu^n\right)
\ol B\sigma^{\mu\nu}\,\frac{\tau^0}{2}B
+
h_{v^nBB}\left(\partial_\mu v_\nu^{na}
-\partial_\nu v_\mu^{na}\right)
\ol B\sigma^{\mu\nu}\,\frac{\tau^a}{2}B
\right)
\label{cub1}
\end{align}
in the four-dimensional baryon-meson effective action.
Here
$\wh v_\mu^n(x)$ and $v^{na}_\mu(x)$ are the $U(1)$ and $SU(2)$ parts
of the $n$th vector meson $v^n_\mu(x)$, respectively, and
$B(x)$ is the baryon field.
Note that we do not include the
direct interaction among baryons and the background gauge 
potential $\cV_\mu^{(+)}$, since the present model exhibits
the complete vector meson dominance as argued in Ref.~\citen{SaSu2}
and emphasized in Ref.~\citen{Hong-Rho-Yee-Yi} for the cases
including baryons.

We recall that the vector and axial-vector mesons couple with
the background gauge potentials as in (\ref{4dim}).
Using these interactions, the Dirac and Pauli form factors are
computed to be
\begin{align}
 \wh F_1(\bk^2)=\sum_{n\ge 1}
\frac{g_{v^n}\wh g_{v^nBB}}{\bk^2+\lambda_{2n-1}} \ ,~~~
 \wh F_2(\bk^2)=\sum_{n\ge 1}
\frac{g_{v^n}\wh h_{v^nBB}}{\bk^2+\lambda_{2n-1}} \ ,~~~
\end{align}
and a similar relation holds for the $SU(2)$ sector.
On the other hand, in the present model, 
(\ref{GhEM}) and (\ref{GEM-2}) together with (\ref{sachs:dp}) yield
the Dirac and Pauli form factors of the form
\begin{align}
 &\wh F_1(\bk^2)=N_c\sum_{n\ge 1}
\frac{g_{v^n}\langle\psi_{2n-1}(Z)\rangle}{\bk^2+\lambda_{2n-1}} \ ,~~~
 \wh F_2(\bk^2)=N_c\left(\frac{g_{I=0}}{2}-1\right)\sum_{n\ge 1}
\frac{g_{v^n}\langle\psi_{2n-1}(Z)\rangle}{\bk^2+\lambda_{2n-1}} \ ,~~~
\nn\\
 &F_1(\bk^2)=\sum_{n\ge 1}
\frac{g_{v^n}\langle\psi_{2n-1}(Z)\rangle}{\bk^2+\lambda_{2n-1}} \ ,~~~
 F_2(\bk^2)=\frac{g_{I=1}}{2}\sum_{n\ge 1}
\frac{g_{v^n}\langle\psi_{2n-1}(Z)\rangle}{\bk^2+\lambda_{2n-1}} \ .~~~
\end{align}
Here we have kept only the leading terms in the large
$N_c$ and large $\lambda$ limit.\footnote{
Since the leading contribution
to the baryon mass $m_B$ is $M_0=8\pi^2\kappa$, we consider $m_B$ to be
of order $\lambda N_c$. However, we will not use the relation
$m_B\simeq M_0$, since we know that the subleading contributions
in $m_B$ are not small as discussed in Ref.~\citen{HSSY}.
}

By comparing these results, we obtain 
\begin{align}
 \wh g_{v^nBB}&=\,
N_c\langle\psi_{2n-1}(Z)\rangle \ ,~~~
 \wh h_{v^nBB}=
N_c\left(\frac{g_{I=0}}{2}-1\right)\,
\langle\psi_{2n-1}(Z)\rangle \ ,~~~
\nn\\
 g_{v^nBB}&=
\langle\psi_{2n-1}(Z)\rangle \ ,~~~
 h_{v^nBB}=
\frac{g_{I=1}}{2}
\,\langle\psi_{2n-1}(Z)\rangle \ .
\label{cub:vec}
\end{align}
It is interesting to note the relation
\begin{align}
\wh g_{v^nBB}=N_c\,g_{v^nBB} \ .
\end{align}
In particular, for the case with $n=1$ and
 $B$ being a nucleon, this means the $g_{\omega NN}=N_c\, g_{\rho NN}$,
which coincides with the constituent quark model
prediction. This relation is found also in Ref.~\citen{Hong-Rho-Yee-Yi}.

The axial form factors $\wh g_A$ and $g_A$ can be obtained
from the effective cubic couplings
with pion and axial-vector mesons of the form
\begin{align}
  \cL_{\rm int}^a=&\,\sum_{n\ge 1}\left(
\wh g_{a^nBB}\,
\wh a_\mu^n\,\ol Bi\gamma_5\gamma^\mu\frac{\tau^0}{2}B
+
g_{a^nBB}\,
a_\mu^{na}\,\ol Bi\gamma_5\gamma^\mu\frac{\tau^a}{2}B
\right) 
\nn\\
&+2i\left(
\wh g_{\pi BB}\,
\wh \pi\,\ol B\gamma_5\frac{\tau^0}{2}B
+
g_{\pi BB}\,
\pi^a\,\ol B\gamma_5\frac{\tau^a}{2}B
\right) \ ,
\label{cub2}
\end{align}
where
$\wh a_\mu^n(x)$ and $a^{na}_\mu(x)$ are the $U(1)$ and $SU(2)$ parts
of the $n$th axial vector meson $a^n_\mu(x)$, respectively, and
$\wh \pi(x)$ and $\pi^{a}(x)$ are the $U(1)$ and $SU(2)$ parts
of the pion field $\Pi(x)$, respectively.
As in (\ref{cub1}), no direct coupling between the baryons and
$\cV^{(-)}_\mu$ is assumed.
It follows from this and (\ref{4dim}) that the invariant amplitude
with the incoming and outgoing baryons in the presence of
the external gauge field $\cV^{(-)}_\mu$ is given by\footnote{The 
propagator of $a_\mu^n$ is given by
that of a Proca field:
$\frac{1}{k^2+\lambda_{2n}}(\eta_{\mu\nu}+{k_\mu k_\nu}/\lambda_{2n})
$.}
\begin{align}
\sqrt{2p^0}\sqrt{2p'^0}\,\ol u(\bp',s')
\frac{\delta_{I_3'I_3}}{2}\Bigg[&
i\gamma_5\gamma_\mu\sum_{n\ge 1}
\frac{g_{a^n}\wh g_{a^nBB}}{k^2+\lambda_{2n}}
\nn\\
&\hspace{-5mm}+k_\mu\gamma_5\left(
\frac{2f_\pi\wh g_{\pi BB}}{k^2}
-2m_B\sum_{n\ge 1}\frac{g_{a^n}\wh g_{a^n BB}}{\lambda_{2n}}
\frac{1}{k^2+\lambda_{2n}}
\right)
\Bigg]u(\bp,s)
\nn\\
&\hspace{-2cm}+
(SU(2)\mbox{ part}) \ .
\end{align}
By comparing this with the matrix elements (\ref{AA0})--(\ref{gAgP}),
we obtain $\wh g_A$, $\wh h_A$, $g_A$, and $h_A$ as functions of
the cubic coupling constants in (\ref{cub2}):
\begin{align}
& \wh g_A(k^2)=\sum_{n\ge 1}
\frac{g_{a^n}\wh g_{a^nBB}}{k^2+\lambda_{2n}} \ ,
\quad
 g_A(k^2)=\sum_{n\ge 1}
\frac{g_{a^n}g_{a^nBB}}{k^2+\lambda_{2n}} \ ,
\end{align}
\begin{align}
& \wh g_P(k^2)=2m_B
\frac{2f_\pi\wh g_{\pi BB}}{k^2}
-4m_B^2\sum_{n\ge 1}\frac{g_{a^n}\wh g_{a^nBB}}{\lambda_{2n}}
\frac{1}{k^2+\lambda_{2n}}\ ,
\nn\\
& g_P(k^2)=2m_B
\frac{2f_\pi g_{\pi BB}}{k^2}
-4m_B^2\sum_{n\ge 1}\frac{g_{a^n}g_{a^nBB}}{\lambda_{2n}}
\frac{1}{k^2+\lambda_{2n}}\ .
\end{align}
Equating these results with (\ref{gA})
then using (\ref{h-g}) and (\ref{sumga2})
leads to the relations
\begin{align}
&\wh g_{a^nBB}
=\frac{N_c}{32\pi^2\kappa}\langle\partial_Z\psi_{2n}(Z)\rangle \ ,
\quad
g_{a^nBB}=\frac{8\pi^2\kappa}{3}\,
\langle\rho^2\rangle\langle\partial_Z\psi_{2n}(Z)\rangle 
\ ,
\\
\nn\\
&\wh g_{\pi BB}=\frac{m_B}{f_\pi}\frac{N_c}{16\pi^3\kappa}
\vev{\frac{1}{k(Z)}}
\ ,
\quad
g_{\pi BB}=\frac{m_B}{f_\pi}\frac{16\pi\kappa}{3}
 \vev{\frac{\rho^2}{k(Z)}}
\ .
\label{pBB}
\end{align}
Comparing (\ref{pBB}) with (\ref{gA0}),
we note that the following relations hold:
\begin{align}
 \wh g_A(0)=\frac{f_\pi\wh g_{\pi BB}}{m_B} \ ,~~~~
 g_A(0)=\frac{f_\pi g_{\pi BB}}{m_B} \ ,
\label{GT2}
\end{align}
i.e., the Goldberger-Treiman relation.

\subsection{Numerical estimate}

By solving (\ref{psi1}) numerically
using the shooting method, the expectation values
of $\psi_{2n-1}(Z)$
and $\partial_Z\psi_{2n}(Z)$ with respect to the wavefunction
$\psi_Z(Z)$ (see Appendix \ref{ap:wv} for details)
can be estimated as\footnote{Here we show values 
only up to $n=4$, as the
quantum treatment of the values 
seems to break down for higher $n$ because its
deviation from the classical value becomes significant.}
\begin{align}
\begin{array}{c||c|cc|cc}
n & \lambda_n 
& \langle\psi_n(Z)\rangle_{n_z=0} 
& \langle\psi_n(Z)\rangle_{n_z=1}
& \langle\partial_Z\psi_n(Z)\rangle_{n_z=0}
& \langle\partial_Z\psi_n(Z)\rangle_{n_z=1}
\\
\hline
1 & 0.669 & 5.80   & 4.51  &  0    & 0      \\
2 & 1.57  & 0      & 0     & 3.46  & 0.618  \\
3 & 2.87  & -2.70  & 0.766 & 0     & 0      \\
4 & 4.54  & 0      & 0     & -3.08  & 2.22   
\end{array}
\end{align}
As an example, consider the nucleon with $n_\rho=n_z=0$.
Using this table, we obtain
\begin{align}
 \begin{array}{c||cc|cc}
n & \wh g_{v^nNN} & g_{v^nNN} & \wh g_{a^nNN} & g_{a^nNN}
\\
\hline
1 &   17.4  & 5.80  & 4.42 &6.14   \\
2 &  -8.10 & -2.70 & -3.84 & -5.46  
 \end{array}
\end{align}
In particular, for the $\rho$ meson we thus obtain
\begin{align}
g_{\rho NN}=g_{v^1NN}\simeq 5.80 \ .
\label{rhoNN}
\end{align}
This is consistent with the experimental data 
$g_{\rho NN}|_{\rm exp}=4.2 - 6.5$ \cite{exp:rhonn}
(see also Ref.~\citen{Hong-Rho-Yee-Yi}).
Note that the functional form of (\ref{cub:vec})
shows the universality of the
$\rho$ meson couplings among the spin 1/2 baryons with
any $n_\rho$ and $n_z=0$.
Moreover, for large $N_c$ and large $\lambda$, 
$g_{v^1BB}$ tends to take a common value for any baryon state $B$
with $I=J=1/2$, 
because the corresponding wavefunction has a narrow support with
a width of $\cO(\lambda^{-1/2}N_c^{-1/2})$.
{}For reference, $g_{\rho N(1535)N(1535)}$, the cubic coupling 
of $\rho$ with $N(1535)$, 
can be numerically 
computed using the quantum wavefunction for $n_z=1$ as
\begin{align}
 g_{\rho N(1535)N(1535)}=\langle\psi_1(Z)\rangle_{n_z=1}
\simeq 4.51 \ .
\end{align}
To see if the $\rho$ meson 
universality holds extensively in the meson sector,
we list
the meson cubic couplings $g_{\rho\pi\pi}$, $g_{\rho v^n v^n}$, and
$g_{\rho a^n a^n}$
computed in Ref.~\citen{SaSu2}:
\begin{align}
g_{\rho\pi\pi}=4.81 ~~~~~~
\begin{array}{c||cc}
n & g_{\rho v^nv^n} &g_{\rho a^na^n} \\
\hline
1 & 5.19 & 3.32 \\
2 & 3.12 & 2.98 \\
3 & 2.93 & 2.89 \\
4 & 2.87 & 2.85
\end{array}
\end{align}
It seems that the
relation $g_{\rho\pi\pi}\sim g_{\rho\rho\rho}\sim
g_{\rho NN}$ roughly holds within 20\% error, although
 $g_{\rho v^n v^n}$ with $n>1$ and $g_{\rho a^na^n}$
are not close enough
to ensure the universality.

If we use $m_B=940~{\rm MeV}$ as an input, together with
(\ref{Mkkkappa}), 
the Yukawa couplings of the pion and the nucleons
given in (\ref{pBB}) are evaluated as\footnote{
$\wh g_{\pi NN}$ is related to the Yukawa coupling
of the isoscalar mesons $g_{\eta NN}$ and $g_{\eta' NN}$.
It is, however, difficult to tell which component
it corresponds to, since we are analyzing massless QCD with $N_f=2$.
}
\begin{align}
 \wh g_{\pi NN}\simeq 5.37 \ ,~~~~
 g_{\pi NN}\simeq 7.46 \ ,
\end{align}
while the experimental value is $g_{\pi NN}|_{\rm exp}\simeq 13.2$
\cite{Bu}.
The smallness of the predicted value is related to
the similar observation for the axial coupling $g_A$ analyzed in
\S \ref{axialcoupling} through the Goldberger-Treiman
relation (\ref{GT2}).
In the same manner, the Yukawa couplings involving
$N(1440)$ and $N(1535)$ are easy to evaluate:
\begin{align}
& \wh g_{\pi N(1440)N(1440)}\simeq 8.23 \ ,~~~~ 
g_{\pi N(1440)N(1440)}\simeq 16.7 \ ,
\nn\\
& \wh g_{\pi N(1535)N(1535)}\simeq 4.55 \ ,~~~~ 
g_{\pi N(1535)N(1535)}\simeq 6.32 \ .
\end{align}
Here we have used (\ref{rho^2-1}), (\ref{rho^2-2}), (\ref{kz-1}), and
(\ref{kz-2}).\footnote{
A computation of the Yukawa couplings involving the excited
baryons is also performed in Ref.~\citen{HoInYe},
where a five-dimensional spinor field 
is incorporated
in the gravity side in the bottom-up approach in holographic QCD.
}

\vspace {1cm}
\section{Summary and discussion}

In this paper, using the model of holographic QCD 
proposed in Refs.~\citen{SaSu1} and 
\citen{SaSu2}, 
we calculated the static quantities
of the nucleon and excited baryons
such as 
$N(1440)$, $N(1535)$, and $\Delta$, and
the form factors of the spin 1/2 baryons.
The baryons are described as quantized instantons in
five-dimensional YM-CS theory \cite{HSSY}. By defining the chiral currents
properly at the spatial infinity in the fifth dimension and by solving the
YM-CS equations of motion using the instanton profile given in
Ref.~\citen{HSSY} and Green's functions, we obtain an explicit expression for
the chiral currents depending on the baryon state. From the currents we
computed various static quantities of the proton/neutron, such as the charge
radii, magnetic moments, axial coupling and axial radius.
See the summary table below. 
The Goldberger-Treiman relation is
naturally derived. These quantities can be computed for excited baryons
in the same manner, which are our theoretical
prediction for the excited baryons
(see the second table below).
We also calculated
the nucleon form factors (and also those for excited
baryons). 
It was shown that the electric and magnetic form factors of 
the nucleon
are roughly consistent with the dipole behavior observed in 
experiments. 
The electric as well as the magnetic charge radii of the baryons and 
their couplings to mesons are 
calculated from the form factors.

A table summarizing our results 
for the static properties for the proton and neutron is
given below. For a comparison, the table includes the values obtained in
experiments, and also the results obtained from the Skyrmion \cite{ANW}.
\vspace{1cm}
\begin{align}
 \begin{array}{c||ccc}
& \mbox{our model} & \mbox{Skyrmion \cite{ANW} } 
& \mbox{experiment} \\
\hline
\vev{r^2}_{I=0}^{1/2} & 0.742\mbox{ fm} &
0.59\mbox{ fm} & 0.806\mbox{ fm} \\
\vev{r^2}_{M,\,I=0}^{1/2} & 0.742\mbox{ fm} &0.92\mbox{ fm} &
0.814\mbox{ fm} \\
\vev{r^2}_{E,\rm p} & (0.742\mbox{ fm})^2 & \infty &
(0.875\mbox{ fm})^2 \\
\vev{r^2}_{E,\rm n} & 0 & -\infty &
-0.116\mbox{ fm}^2 \\
\vev{r^2}_{M,\rm p} & (0.742\mbox{ fm})^2 & \infty &
(0.855\mbox{ fm})^2 \\
\vev{r^2}_{M,\rm n} & (0.742\mbox{ fm})^2 & \infty &
(0.873\mbox{ fm})^2 \\
\vev{r^2}_A^{1/2}& 0.537\mbox{ fm} & - & 0.674\mbox{ fm}
\\
\mu_p & 2.18 & 1.87 & 2.79 \\
\mu_n & -1.34 & -1.31 & -1.91 \\
\left|\frac{\mu_p}{\mu_n}\right|& 1.63 & 1.43 & 1.46\\
g_A& 0.734 & 0.61 & 1.27
\\
g_{\pi NN} & 7.46 & 8.9 & 13.2
\\
g_{\rho NN} & 5.80 & - & 4.2\sim 6.5
 \end{array}
\nonumber
\end{align}
\vspace{1cm}

{}For excited baryons, $N(1440)$ (Roper) and $N(1535)$,
we provide our theoretical predictions
in the following table.\newpage
\begin{align}
 \begin{array}{c||ccc}
& n,p & N(1440) & N(1535) \\
\hline
\vev{r^2}_{E,\rm p}
 & (0.742\mbox{ fm})^2 & (0.742\mbox{ fm})^2 & 
(0.699\mbox{ fm})^2  \\
\vev{r^2}_{E,\rm n} & 0 & 0 &
0 \\
\vev{r^2}_{M,\rm p}
 & (0.742\mbox{ fm})^2 & (0.742\mbox{ fm})^2 & 
(0.699\mbox{ fm})^2  \\
\vev{r^2}_{M,\rm n}
 & (0.742\mbox{ fm})^2 & (0.742\mbox{ fm})^2 & 
(0.699\mbox{ fm})^2  \\
\vev{r^2}_A^{1/2}& 0.537\mbox{ fm} & 0.537\mbox{ fm}
 & 0.435\mbox{ fm} \\
\mu_p & 2.18 & 2.99 & 2.18 \\
\mu_n & -1.34 & -2.15 & -1.34 \\
\left|\frac{\mu_p}{\mu_n}\right|& 1.63 & 1.39 & 1.63
\\
g_A& 0.734 & 1.07 & 0.380\\
g_{\pi BB} & 7.46 & 16.7 & 6.32
\\
g_{\rho BB} & 5.80 & 5.80 & 4.51
 \end{array}
\nonumber
\end{align}

As shown in the first 
table, we found good agreement with experiments
for various quantities of baryons.
Our numerical results presented here should be treated with
caution since they were obtained for large values of the 't Hooft coupling and
$N_c$. In order to incorporate the difference among 
excited baryon states,
we included subleading corrections only at the last stage of the
calculations. This procedure is difficult to justify since we
considered the leading order action (\ref{model}) as our starting point,
and there should be more corrections.
Further investigation to improve the accuracy of the results
would be interesting. 

In this paper, baryons have been described as a soliton,
and the baryon physics has been analyzed by
a standard semiclassical quantization of the soliton.
This approach appears to be completely different from
 the treatment of Ref.~\citen{Hong-Rho-Yee-Yi},
in which a key step is to introduce a five-dimensional
spinor field into the five-dimensional YM-CS system
to represent the baryons.
In order to relate the two approaches, consider
the $\wh vBB$ coupling, for simplicity,
in the four-dimensional
meson-baryon effective Lagrangian (\ref{cub1}):
\begin{align}
\cL_{\rm int}^v=&\sum_{n\ge 1}\left(
\wh g_{v^nBB}\,
\wh v_\mu^n(x)\,\ol B(x)i\gamma^\mu\frac{\tau^0}{2}B(x)
+\cdots \right)\ .
\end{align}
If we substitute $\wh g_{v^nBB}$ from (\ref{cub:vec}),
this can be written as
\begin{align}
 \cL_{\rm int}^v=\,\frac{N_c}{2}\int dz\,\left(
\wh A_\mu(x,z)\,
\ol\cB(x,z)i\gamma^\mu\frac{\tau^0}{2}\cB(x,z)
+\cdots \right)\ .
\end{align}
Here $\wh A_\mu$ is the $U(1)$ part of
the five-dimensional gauge potential in (\ref{exp}), and
$\cB(x,z)\equiv B(x)\psi_Z(z)$ is a five-dimensional spinor
field constructed by the four-dimensional baryon field $B(x)$ and
the suitably normalized wavefunction $\psi_Z(Z)$ corresponding
to the baryon $B$.
This term can be summarized as a five-dimensional gauge interaction 
by regarding $\psi_Z(Z)$ as an eigenmode of a wave equation
that reproduces the baryon spectrum found in Ref.~\citen{HSSY}.
Likewise, the interaction terms given in (\ref{cub1}) and 
(\ref{cub2}) are sufficient to reconstruct the 
five-dimensional spinor field action
with the Pauli-type interaction in Ref.~\citen{Hong-Rho-Yee-Yi}.

We have concentrated on a one-point function of mesons and
the electromagnetic field in the presence of a single quantized 
soliton, mainly to extract the static properties of baryons. 
There are other interesting
aspects of the force associated with the baryons, for example, 
interactions between baryons, in particular. 
When baryons are far from each other, we
can use our results and compute the nuclear force based on
a one-meson-exchange picture of the nuclear force. Resolving the issue
of the nuclear force at short distance is interesting and will be
reported in our forthcoming paper \cite{HSS2}.

\section*{Acknowledgements}

We would like to thank M.~Harada, H.~Hata, D. K.~Hong,
K.~Itakura, M.~Murata, H.~Murayama, S.-J.~Rey, M.~Rho, S.~Yamato,
H.-U.~Yee and P.~Yi for discussions.
We would also like to thank the Yukawa Institute of Theoretical Physics at
Kyoto University and the Galileo Galilei Institute for Theoretical
Physics, where part of this work was done, for their kind hospitality
and the INFN for partial support during the completion of this work.
The work of K.H. and T.S. was partly supported by 
a Grant-in-Aid for Young Scientists (B), from the Ministry of Education, 
Culture, Sports, Science and Technology (MEXT), Japan.
The work of S.S. was supported in part by a
JSPS Grant-in-Aid for Creative Scientific Research
No. 19GS0219 and also by the World Premier International
Research Center Initiative (WPI Initiative), MEXT, Japan.

\appendix

\section{Excited Baryons}

In this appendix, first we summarize quantum wavefunctions for
the
lowest/excited states of baryons obtained in Ref.~\citen{HSSY}, and present
detailed calculations relevant to the static quantities of the baryons 
presented in \S \ref{secstatic}.

\subsection{Wavefunctions for  excited baryons }
\label{ap:wv}

In Ref.~\citen{HSSY}, fluctuations around the classical solution
(\ref{HSSYsol}) of the five-dimensional YM-CS theory
in the curved background (\ref{model}) are
quantized.
The quantization of the soliton was done in the approximation of 
slowly moving (pseudo)moduli of the soliton solution. The moduli
degrees of freedom of the soliton are
\begin{eqnarray}
 X^i(t)\  ,~~~ Z(t)\  ,~~~ \rho(t)\  ,~~~a^I(t)\  .
\end{eqnarray}
$X^i$ and $Z$ describe the center-of-mass motion of the
soliton, while $\rho$ and $a^I$ ($I=1,2,3,4$) describe the
size of the soliton, which is instanton-like in the $SU(2)$ sector,
and the orientation of the instanton in the $SU(2)$ group space,
respectively, with $(a^I)^2=1$. 
The Hamiltonian for the moduli (\ref{Ham}) provides the quantized energy
eigenstates of the baryon, specified by the quantum number
$B=(l,I_3,n_\rho, n_z)$ and its spin $s$.
The 
wavefunctions for the quantized states can be
written explicitly for low-lying states. The nucleon wavefunctions,
$B=(1,\pm 1/2,0,0)$ and $s=1/2$, are written as (\ref{wf}), 
\begin{eqnarray}
 \ket{\,p\uparrow\,}\propto
R(\rho)\psi_Z(Z)(a_1+ia_2)\ ,~~~
 \ket{\,n\uparrow\,}\propto
R(\rho)\psi_Z(Z)(a_4+ia_3)\ ,
\end{eqnarray}
with 
\begin{eqnarray}
 R(\rho)=\rho^{-1+2\sqrt{1+N_c^2/5}}
e^{-\frac{M_0}{\sqrt{6}}\rho^2}\ ,~~
\psi_Z(Z)=e^{-\frac{M_0}{\sqrt{6}}Z^2}\ .
\label{Rnuc}
\end{eqnarray}

Here we present the first excited states for which static
quantities are computed in \S \ref{secstatic}.
The excited state with $B=(1,\pm 1/2,1,0)$ corresponds to $N(1440)$,
which is called the Roper excitation. The wavefunction is
\begin{eqnarray}
 R(\rho) = 
\left(
\frac{2M_0}{\sqrt{6}}\rho^2-1-2\sqrt{1+N_c^2/5}
\right)
\rho^{-1+2\sqrt{1+N_c^2/5}}
e^{-\frac{M_0}{\sqrt{6}}\rho^2}\ .~~
\label{roperwf}
\end{eqnarray}
The excited state with $B=(1,\pm 1/2,0,1)$ is $N(1535)$. 
The wavefunction for the $Z$ part is now given as
\begin{eqnarray}
 \psi_Z(Z)=Z e^{-\frac{M_0}{\sqrt{6}}Z^2}\ ,
\label{npwf}
\end{eqnarray}
where, again, its normalization 
constant is not fixed yet. The other part of the wavefunction 
is the same as that of the proton/neutron. This excitation has
a negative parity, which is reflected in the oddness of the 
wavefunction $\psi_Z(Z)$.

The excited states in the $SU(2)$ group space have already been studied in the
context of the Skyrmion \cite{ANW}. 
The lightest among these excited states is
$\Delta$ with $I=J=3/2$. The isoquartet is composed of
the four baryons $\Delta^{++},\Delta^+, \Delta^0$, and $\Delta^-$
with $I_3=3/2,1/2,-1/2$, and $-3/2$, respectively.
As shown in Ref.~\citen{ANW}, $\Delta^{++}$ with 
$s=3/2$ is described by the wavefunction
\begin{eqnarray}
 (a_1 + ia_2)^3\  .
\label{delta++}
\end{eqnarray}
The rest of the $\Delta$ wavefunctions with $s=3/2$ are
obtained by letting the
isospin lowering operator act on (\ref{delta++}):
\begin{align}
& (a_1+ia_2)^2 (a_4+ia_3)\  ,~~~(\mbox{for }\Delta^+) \nn\\
& (a_1+ia_2)(a_4+ia_3)^2\  ,~~~(\mbox{for }\Delta^0) \nn\\
& (a_4+ia_3)^3\  .~~~(\mbox{for }\Delta^-) \ 
\label{delta+}
\end{align}
The wavefunction $R(\rho)$ for $\Delta$ (states with $l=3$ and
$n_\rho=0$) is given as
\begin{equation}
 R(\rho)=\rho^{-1+2\sqrt{4+N_c^2/5}}
e^{-\frac{M_0}{\sqrt{6}}\rho^2}\ .
\label{Rdel}
\end{equation}

We use these wavefunctions to compute the static quantities of excited
baryons.

\subsection{Expectation values relevant to $N(1440)$ and $N(1535)$}
\label{xx}

First, we summarize what quantities are necessary for computing various static
quantities presented in \S \ref{secstatic}:
\begin{eqnarray}
 \langle F(Z)\rangle\  , \quad 
 \langle \rho^2\rangle\  , \quad 
 \left\langle \frac{\rho^2}{k(Z)}\right\rangle\  , \quad 
 \langle F_A(Z)\rangle\  .
\end{eqnarray}
The first quantity $F(Z)$ is necessary for computing 
$\langle r^2\rangle_{I=0}$, and thus
$\langle r^2\rangle_{E,M}$.
The second, third, and fourth quantities,  
are for the magnetic moment $\mu$, the axial coupling $g_A$, and 
the axial radius $\langle r^2\rangle_{A}^{1/2}$, respectively.

First, we compute $\langle \rho^2\rangle$. 
The integral to be evaluated is
\begin{eqnarray}
 \langle \rho^2 \rangle_{n_\rho=1} = 
\frac{\int \rho^5 R(\rho)^2 d\rho}{\int \rho^3 R(\rho)^2 d\rho}\  ,
\end{eqnarray}
and we substitute (\ref{roperwf}) into this to allow for the Roper excitation.
By partial integration, we obtain
\begin{eqnarray}
&&   \langle \rho^2 \rangle_{n_\rho=0} 
= \rho_{\rm cl}^2 \frac{\sqrt{5}+ 2\sqrt{5 + N_c^2}}{2 N_c} \ ,
\\
&&   \langle \rho^2 \rangle_{n_\rho=1} 
= \rho_{\rm cl}^2 \frac{3\sqrt{5}+ 2\sqrt{5 + N_c^2}}{2 N_c} \ .
\end{eqnarray}
The former equation is for the proton/neutron, (\ref{vevrho}).
For $N_c=3$, these are numerically given by 
\begin{eqnarray}
&&   \langle \rho^2 \rangle_{n_\rho=0} 
\simeq 1.62 \times \rho_{\rm cl}^2\  , 
\label{rho^2-1}
\\
&&   \langle \rho^2 \rangle_{n_\rho=1} 
\simeq 2.37 \times 
\rho_{\rm cl}^2\  . 
\label{rho^2-2}
\end{eqnarray}

Next, we compute 
$\left\langle \frac{1}{k(Z)}\right\rangle$,
$\langle F(Z)\rangle$, and $\langle f_A(Z)\rangle$, that is,
the quantities relevant to the $Z$
directions. This is necessary for the $N(1535)$ excitation.
The classical values are given with $Z=0$ as
\begin{eqnarray}
\left\langle \frac{1}{k(Z)}\right\rangle
\simeq 1\ ,~~~
\quad 
\langle F(Z)\rangle
\simeq F(0)\ ,~~~
\quad
\langle F_A(Z)\rangle
\simeq F_A(0)\  .
\end{eqnarray}
In the large $\lambda$ expansion, the wavefunction for $Z$ is localized
at the origin $Z=0$. Inclusion of the subleading terms causes
differences
in baryon states.

Numerical evaluation using the wavefunction (\ref{npwf}) is performed as
follows: 
\begin{eqnarray}
&&
 \left\langle \frac{1}{k(Z)}\right\rangle_{n_z=0}
= \frac{\int dZ \; \frac{1}{1+Z^2}e^{-2\frac{M_0}{\sqrt{6}}\, Z^2}}
{\int dZ \; e^{-2\frac{M_0}{\sqrt{6}}\, Z^2}}
\simeq 0.649\ ,
\label{kz-1}
\\
&&
 \left\langle \frac{1}{k(Z)}\right\rangle_{n_z=1}
= \frac{\int dZ \; \frac{Z^2}{1+Z^2}e^{-2\frac{M_0}{\sqrt{6}}\, Z^2}}
{\int dZ \; Z^2 e^{-2\frac{M_0}{\sqrt{6}}\, Z^2}}
\simeq 0.337\ ,
\label{kz-2}
\end{eqnarray}
where we have used $\kappa=0.00745$. 

For $F(Z)$ and $F_A(Z)$, we first solve the differential
equations satisfied by them, (\ref{F}) and (\ref{fAd}), 
and then use the solutions to numerically evaluate
the normalized integrals. We obtain
\begin{eqnarray}
&&  \left\langle F(Z)\right\rangle_{n_z=0} \simeq 12.7 \  ,
\quad 
  \left\langle F(Z)\right\rangle_{n_z=1} \simeq 10.9 \  ,
\\
&&  \left\langle F_A(Z)\right\rangle_{n_z=0} \simeq 6.67 \  ,
\quad 
  \left\langle F_A(Z)\right\rangle_{n_z=1} \simeq 4.38\ .
\end{eqnarray}
Classical values are $F(0)=14.3$ and $F_A(0)= 7.82$; thus, the ratios to the
classical values are given as 
\begin{eqnarray}
&&  \frac{\left\langle F(Z)\right\rangle_{n_z=0}}{F(0)} \simeq 0.892 \  ,
\quad 
  \frac{\left\langle F(Z)\right\rangle_{n_z=1}}{F(0)} \simeq 0.762 \  ,
\\
&& \frac{\left\langle F_A(Z)\right\rangle_{n_z=0}}{F_A(0)} \simeq 0.852\  ,
\quad 
  \frac{\left\langle F_A(Z)\right\rangle_{n_z=1}}{F_A(0)} \simeq 0.560 \  .
\end{eqnarray}
Thus, smaller values of the quantities are obtained for $n_z=1$.

Using these expectation values, the static properties of $N(1440)$ and
$N(1535)$ can be computed, as presented in \S \ref{secstatic}.

\subsection{Magnetic moment of $\Delta$}
\label{xx2}

Evaluation of the isoscalar and isovector magnetic moments 
requires the matrix elements of
\begin{eqnarray}
\rho^2 \chi^i = -i \rho^2\tr\left(
\tau^i \boldsymbol{a}^{-1}\dot{\boldsymbol{a}}
\right)\  , 
\quad
\tr \left(\boldsymbol{a} \tau^i \boldsymbol{a}^{-1}\tau^a\right)\  ,
\quad
\rho^2\ .
\end{eqnarray}
Here,
we compute these for the wavefunctions of 
$\Delta$ given in (\ref{delta++}), (\ref{delta+}) and (\ref{Rdel}). 

First of all, the third component of the spin of these states is chosen
to be $s=+3/2$. Therefore, using (\ref{chi}), we can see that the value
of $\rho^2\chi^{i=3}$ is three times that of the proton/neutron for which $s=1/2$.

Next, let us evaluate 
$\tr \left(\boldsymbol{a} \tau^i \boldsymbol{a}^{-1}\tau^a\right)$.
{}From the symmetric nature of this quantity, it is enough to consider
the index $i=a=3$. For this choice of the index, we have
\begin{eqnarray}
 \tr \left(\boldsymbol{a} \tau^i \boldsymbol{a}^{-1}\tau^a\right)
 = 4 a_4^2 + 4 a_3^2 -2 \  .
\end{eqnarray}
In order to calculate the expectation value of this quantity
with the wavefunctions (\ref{delta++}) and (\ref{delta+}), we
use
the spherical coordinates
\begin{eqnarray}
&& a_4 = \cos\theta_0 \ ,\\
&& a_3 = \sin\theta_0 \cos\theta_1 \ , \\
&& a_2 = \sin\theta_0 \sin\theta_1 \cos\theta_2 \ , \\
&& a_1 = \sin\theta_0 \sin\theta_1 \sin\theta_2 \  .
\end{eqnarray}
The Jacobian for this change of variables is
$d\Omega_3 = \sin^2\theta_0 \sin\theta_1
d\theta_0 d\theta_1 d\theta_2$.
Then we obtain, for instance,
\begin{eqnarray}
&& \langle \Delta^{++}| (4 a_4^2 + 4 a_3^2 -2 )|\Delta^{++}\rangle
\nonumber \\
&&
\hspace{5mm}
= 
\frac{
2\pi \int_0^\pi\int_0^\pi 
d\theta_0 d\theta_1
\sin^8\theta_0 \sin^7 \theta_1 
(4 \cos^2\theta_0 + 4 \sin^2\theta_0 \cos^2\theta_1-2)
}
{\int d\Omega_3 (a_1^2 + a_2^2)^3}
\nonumber \\
&&
\hspace{5mm}
= -\frac{6}{5} \  ,  \\[10pt]
&& \langle \Delta^{+}| (4 a_4^2 + 4 a_3^2 -2 )|\Delta^{+}\rangle
\nonumber \\
&&
= 
\frac{
2\pi \int_0^\pi\int_0^\pi
 d\theta_0 d\theta_1
\sin^6\!\theta_0 \sin^5\! \theta_1 
( \cos^2\!\theta_0 \!+\!  \sin^2\!\theta_0 \cos^2\!\theta_1)
(4 \cos^2\!\theta_0 \!+\! 4 \sin^2\!\theta_0 \cos^2\!\theta_1\!-\!2)
}
{\int d\Omega_3 (a_1^2 + a_2^2)^2 (a_4^2+a_3^2)}
\nonumber \\
&& 
= -\frac{2}{5} \  .
\end{eqnarray}
The denominators originate from
the normalization of the states (\ref{delta++}) and
(\ref{delta+}).
It is also easy to see that
\begin{align}
 \langle \Delta^{0}| (4 a_4^2 + 4 a_3^2 -2 )|\Delta^{0}\rangle
 =\frac{2}{5} \ , ~~~
 \langle \Delta^{-}| (4 a_4^2 + 4 a_3^2 -2 )|\Delta^{-}\rangle
 =\frac{6}{5} \ .
\end{align}
The expectation value of $\rho^2$ with respect to the wavefunction
(\ref{Rdel}) is
\begin{align}
 \vev{\rho^2}_{l=3}=
\frac{\sqrt{5}+2\sqrt{20+N_c^2}}{2N_c}\,\rho^2_{\rm cl} \ ,
\label{vevrho-l3}
\end{align}
and the ratio to (\ref{vevrho}) is
\begin{align}
c\equiv\frac{\vev{\rho^2}_{l=3}}{\vev{\rho^2}_{l=1}}
=\frac{\sqrt{5}+2\sqrt{20+N_c^2}}{\sqrt{5}+2\sqrt{5+N_c^2}}\simeq 1.34\ .
\end{align}
for $N_c=3$.

Comparing these with
the proton state with up spin, we obtain 
(\ref{md++}) and (\ref{md+}). These ratios are
used to compute the
magnetic moments of $\Delta$ as
\begin{eqnarray}
&& \mu_{\Delta^{++}}= 3 \times 
\frac12 \left(\mu_p + \mu_n\right)+ \frac{9c}{5} 
\times  \frac12 \left(\mu_p - \mu_n\right) \simeq 5.50\  , \\
&& \mu_{\Delta^{+}}= 3 
\times \frac12 \left(\mu_p + \mu_n\right)+
\frac{3c}{5} \times
\frac12 \left(\mu_p - \mu_n\right)\simeq 2.67 \  ,\\
&& \mu_{\Delta^{0}}= 3 
\times \frac12 \left(\mu_p + \mu_n\right)-
\frac{3c}{5} \times
\frac12 \left(\mu_p - \mu_n\right)\simeq -0.15 \  ,\\
&& \mu_{\Delta^{-}}= 3 \times 
\frac12 \left(\mu_p + \mu_n\right)- \frac{9c}{5} 
\times  \frac12 \left(\mu_p - \mu_n\right) \simeq -2.97\  , 
\end{eqnarray}
i.e., (\ref{deltam}).
Here we have 
substituted
the magnetic moment for the proton and the neutron, 
(\ref{mmpn}).

\section{Useful Formulae and Notation}

\subsection{Useful formulae for currents}

Here we summarize useful formulae that are used in 
\S \ref{seccurrents} for computing the currents.
For $g$ defined in (\ref{defg}), the following equations are obtained, 
\begin{eqnarray}
g\del_i g^{-1}&=&+\frac{i}{\xi^2}
\left(
(z-Z)\tau^i-\epsilon_{ijb}(x^j-X^j)\tau^b
\right)\ ,\\
g\del_z g^{-1}&=&-\frac{i}{\xi^2}(x^b-X^b)\tau^b
\ ,
\\
g^{-1}\del_i g&=&-\frac{i}{\xi^2}
\left(
(z-Z)\tau^i+\epsilon_{ijb}(x^j-X^j)\tau^b
\right)\ ,\\
g^{-1}\del_z g&=&+\frac{i}{\xi^2}(x^b-X^b)\tau^b
\ .
\end{eqnarray}
In particular, its relation to $SU(2)$ generators is given as
\begin{eqnarray}
g\tau^a g^{-1}&=&\frac{1}{\xi^2}\Big[
\left((z-Z)^2-|\vec x-\vec X|^2\right)\tau^a+\nn\\
&&~~+2\left(\epsilon^{bac}(x^b-X^b)(z-Z)+(x^a-X^a)(x^c-X^c)
\right)\tau^c
\Big]
\ ,\quad
\\
\left[\,g^{-1}\del_i g\, ,\,\tau^a\,\right]
&=&\frac{2}{\xi^2}\left(
(x^b-X^b)\delta^{ia}-(x^a-X^a)\delta^{ib}
+(z-Z)\epsilon^{iab}
\right)\tau^b \ ,
\\
\left[\,g^{-1}\del_z g\, ,\, \tau^a\,\right]
&=&-\frac{2}{\xi^2}
(x^k-X^k)\epsilon^{kab}\tau^b \ ,
\\
g\tau^b+\tau^b g&=&\frac{2}{\xi}\left((z-Z)\tau^b-i(x^b-X^b)\right)\ ,\\
g\tau^b-\tau^b g&=&\frac{2}{\xi}(x^a-X^a)\,\epsilon^{abc}\tau^c \ .
\end{eqnarray}
A useful identity for the epsilon tensor is
\begin{eqnarray}
\epsilon_{ija}|\vec x|^2=
\epsilon_{abj}x^ix^b
-\epsilon_{abi}x^jx^b
+\epsilon_{ijb}x^ax^b\ .
\end{eqnarray}

\subsection{Useful formulae for Dirac spinors}
\label{dirac}

We summarize the formulae of the gamma matrix and the Dirac spinor,
which are used in the previous sections.

The Dirac representation of the gamma matrix is taken as
\begin{align}
 \gamma^0=-i\left(
\begin{array}{cc}
1& 0 \\
0&-1
\end{array}
\right) \ ,~~~
 \gamma^j=-i\left(
\begin{array}{cc}
0 & \sigma^j \\
-\sigma^j &0
\end{array}
\right) \ ,~~~
 \gamma_5=i\gamma^0\gamma^1\gamma^2\gamma^3=
\left(
\begin{array}{cc}
0 & 1 \\
1 &0
\end{array}
\right) \ . ~~~
\end{align}
We define
\begin{align}
 \sigma^{\mu\nu}=\frac{i}{2}\left[\gamma^\mu,\gamma^\nu\right] \ .
\end{align}
It can be verified that
\begin{align}
 \gamma^0\gamma^{\mu\dagger}\gamma^0=+\gamma^\mu \ ,~~~
 \gamma^0\gamma_5^\dagger\gamma^0=+\gamma_5 \ ,~~~
 \gamma^0\sigma_{\mu\nu}^\dagger\gamma^0=-\sigma_{\mu\nu} \ .
\end{align}

The on-shell Dirac spinor is given by
\begin{align}
 u(\bp,s)=\frac{1}{\sqrt{2E}}\left(
\begin{array}{c}
\sqrt{E+m_B}\,\chi_{(s)} \\
\sqrt{E-m_B}\,\bn\cdot\bsig\chi_{(s)} \\
\end{array}
\right) \ ,
\end{align}
with
\begin{align}
 \bn=\frac{\bp}{|\bp|} \ ,
\end{align}
\begin{align}
 \chi_{(1/2)}=\left(
\begin{array}{c}
1 \\
0
\end{array}
\right) \ ,~~~
 \chi_{(-1/2)}=\left(
\begin{array}{c}
0 \\
1
\end{array}
\right) \ .
\end{align}
This satisfies the Dirac equation
\begin{align}
 (i\!\psl+m_B)u(\bp,s)=0 \ ,~~~
 \ol u(\bp,s) (i\!\psl+m_B)=0 \ ,
\end{align}
with
\begin{align}
\ol u=u^{\dagger}\beta \ .~~~(\beta=i\gamma^0)
\end{align}
The Dirac spinor is normalized as
\begin{align}
 \ol u(\bp,s')u(\bp,s)=\frac{m_B}{p^0}\,\delta_{ss'} \ .
\end{align}

In the nonrelativistic limit, the Dirac spinor reduces to
\begin{align}
 u(\bp,s)=
\left(
\begin{array}{c}
\chi_{(s)} \\
\frac{1}{2m_B}\,\bp\cdot\bsig\chi_{(s)}
\end{array}
\right)
+\cO(m_B^{-2}) \ .
\end{align}
It is easy to verify 
\begin{align}
 \ol u(\bp',s')\gamma_5\,u(\bp,s)&=
\frac{1}{2m_B}k_a\,(\sigma^a)_{s's}+\cO(m_B^{-2}) \ ,
\\
 \ol u(\bp',s')\,\gamma_5\gamma^0\,u(\bp,s)&
=\frac{i}{2m_B}\,(p+p')_a\,(\sigma^a)_{s's}+\cO(m_B^{-2}) \ , 
\\
 \ol u(\bp',s')\,\gamma_5\gamma^j\,u(\bp,s)&=
i(\sigma^j)_{s's}+\cO(m_B^{-2}) \ .
\end{align}
Here $k=p-p'$ and we used
\begin{align}
 \chi_{(s')}^\dagger\,\sigma^a\,\chi_{(s)}=(\sigma^a)_{s's} \ .
\end{align}

\end{document}